\documentclass[journal]{IEEEtran}
\usepackage[utf8]{inputenc}
\usepackage[T1]{fontenc}

\usepackage{graphicx}
\usepackage{amsmath,amssymb,amsfonts,mathtools}
\usepackage{latexsym}
\usepackage{dsfont}
\usepackage{url}
\usepackage{enumerate}
\usepackage{color}
\usepackage{comment}
\usepackage{cite}
\usepackage{stfloats}
\usepackage{array}

\usepackage{algorithm}
\usepackage{algpseudocode}

\usepackage{tikz}
\usetikzlibrary{arrows.meta,positioning,fit,calc,decorations.pathreplacing}

\newtheorem{proposition}{Proposition}

\ifCLASSINFOpdf
\else
\fi

\definecolor{blue}{rgb}{0,0,0}
\begin{document}

	\title{AirPASS: Over-the-Air Federated Learning via Pinching Antenna Systems}
	
	\author{{Seyed Mohammad Azimi-Abarghouyi,~\IEEEmembership{Member,~IEEE} and Christopher G. Brinton,~\IEEEmembership{Senior Member,~IEEE}}

		\thanks{S. M. Azimi-Abarghouyi is with the Department of Electrical Engineering, Chalmers University of Technology, Gothenburg, Sweden (Email: azimimo@chalmers.se). C. G. Brinton is with the School of Electrical and Computer Engineering, Purdue University, West Lafayette, IN USA (Email: cgb@purdue.edu).}
	}
	
	\maketitle
	
	\vspace{-5pt}

	\begin{abstract}
		This paper investigates over-the-air federated learning (AirFL) in wireless systems where the access point is equipped with a multi-waveguide pinching antenna system (PASS). We adopt the widely studied learning-oriented AirFL formulation, which seeks to maximize the number of selected devices while keeping the aggregation distortion below a prescribed threshold. The resulting joint optimization of device selection, receive beamforming, and pinching-antenna placement is highly nonconvex due to the intricate coupling among these system variables. To address this challenge, we develop AirPASS, an alternating optimization framework with two main components: a homotopy-Riemannian margin-consolidation method for device selection and receive beamforming under \textcolor{blue}{a} fixed PASS configuration, and a homotopy-assisted geometry optimization method for updating the pinching-antenna positions under fixed selected devices and beamformer. \textcolor{blue}{Experiments show that, with the same number of radio-frequency (RF) chains, AirPASS substantially increases device participation and improves learning accuracy over co-located multiple-input multiple-output (MIMO) and fixed-position PASS receivers, and attains an accuracy comparable to that of a co-located array with one RF chain per pinching antenna. Moreover, it matches the participation of semidefinite-relaxation-based selection to within $1$~dB in signal-to-noise ratio (SNR) at a small fraction of its runtime, and it clearly outperforms matching-pursuit scheduling.}
	\end{abstract}
	
	\vspace{-5pt}
	
	\section{Introduction}
	
	Federated learning (FL) enables multiple devices to collaboratively train a shared machine learning model while keeping their raw data local \cite{McMahan}. Devices perform local updates using private datasets and periodically transmit model updates to a central server for aggregation into a global model. This paradigm is particularly attractive for wireless networks, where data is generated at the edge and privacy concerns often limit centralized data collection. However, deploying FL over wireless networks introduces fundamental communication challenges. Modern learning models typically contain millions of parameters, and exchanging model updates over many training rounds can incur significant communication latency and bandwidth overhead. These limitations motivate the design of communication-efficient learning mechanisms that tightly integrate wireless communication and distributed training.
	
	A promising solution is \emph{over-the-air federated learning} (AirFL), which builds upon the principle of \emph{over-the-air computation} (AirComp) \cite{airflmag}. AirComp exploits the waveform superposition property of wireless multiple-access channels to compute functions of distributed signals directly in the air \cite{aircomp}. In AirFL systems, devices simultaneously transmit analog-modulated model updates, and the access point (AP) directly receives their superposition, which corresponds to the aggregated model update required for FL. By integrating communication and aggregation at the physical layer, AirFL significantly reduces communication latency compared to conventional orthogonal multiple access aggregation schemes. \textcolor{blue}{Alternative ways of exploiting wireless superposition, such as \emph{out-of-air computation} (AirCPU) \cite{aircpu}, have also recently emerged.}
	
	Recent years have witnessed extensive research on AirFL system design. Early works studied FL over wireless fading channels and demonstrated the advantages of analog aggregation \cite{FLFadingChannels}. Subsequent studies investigated AirFL designs that jointly optimize wireless transmission and learning performance. In particular, receive beamforming and device selection have been widely studied \cite{DCSDR, BeamformingVectorDesignFL,DeviceSchedulingOTAFL}. Existing approaches mainly fall into two categories. One line of work relies on semidefinite-relaxation (SDR) with difference-of-convex (DC) formulations, referred to hereafter as SDR-DC methods \cite{DCSDR,BeamformingVectorDesignFL}, which are algorithmically powerful but often computationally expensive due to matrix lifting. Another line adopts greedy or matching-pursuit strategies \cite{DeviceSchedulingOTAFL}, which are computationally lighter but may suffer from irreversible early scheduling decisions. Other works extended AirFL by considering weighted aggregation \cite{WeightedAggregationFL}, adaptive transmission strategies \cite{AdaptiveOTAFL}, digital modulation \cite{fedcpu}, and hierarchical architectures for scalable learning across large networks \cite{HierarchicalOTAFL}. In addition, programmable wireless environments such as intelligent reflecting surfaces (IRS) have been explored to enhance performance \cite{IRSFL,IRSFLBroadband}, while integrated sensing and communication (ISAC) has been considered to enable multi-functional operation \cite{isacfl}. 
	
	A key characteristic of most existing AirFL works is the adoption of a \emph{learning-oriented system design} \cite{airflmag}. In particular, the widely studied formulation aims to \emph{maximize the number of selected devices} while ensuring that \textit{the AirComp aggregation distortion remains below a prescribed threshold} \cite{DCSDR,BeamformingVectorDesignFL,DeviceSchedulingOTAFL, IRSFL, isacfl}. This formulation reflects the fundamental learning objective of FL: incorporating updates from a maximal number of devices improves statistical efficiency and model generalization. \textcolor{blue}{At the same time}, the communication design must tightly constrain aggregation distortion to guarantee learning convergence  \cite{airflmag}.
	
	Concurrently, \emph{pinching antenna systems} (PASS) have recently emerged as a novel wireless architecture, offering flexible spatial radiation control through movable antenna elements. In PASS architectures, electromagnetic waves propagate along dielectric waveguides and movable pinching antennas extract and radiate energy at adjustable locations along the waveguide. \textcolor{blue}{Unlike conventional antenna arrays with fixed element positions, PASS can thus dynamically reshape the effective array, both by placing radiating points closer to the devices and by using wavelength-scale displacements to adjust the phases with which their signals combine.} 
	
	The fundamental principles and architectures of PASS were established in \cite{PASSTutorial,PASSPrinciples,PASSArchitecture}. A growing body of work has since explored PASS-based communication systems, including array gain characterization \cite{ArrayGainPASS}, power radiation modeling and beamforming design \cite{PASSPowerRadiation}, performance analysis \cite{PerformancePASS}, \textcolor{blue}{multi-user transmission structures and multiple access \cite{PASSMultiuser,WDMAPASS}}, extensions to multi-waveguide systems \cite{PASSActivation, MIMOPASS}, and ISAC applications \cite{PASSISAC1,PASSISAC2}. Collectively, these results highlight the significant design flexibility of PASS compared with conventional antenna arrays, while preserving relatively low hardware complexity. \textcolor{blue}{Multi-user designs are particularly relevant to the multi-waveguide architecture considered in this paper. For downlink multi-user communications, waveguide multiplexing, waveguide division, and waveguide switching transmission structures have been proposed and combined with joint baseband and pinching beamforming design under max-min fairness \cite{PASSMultiuser}, while waveguide division multiple access (WDMA) assigns a dedicated waveguide to each user and exploits pinching beamforming to enhance the intended signal and suppress inter-user interference \cite{WDMAPASS}. In these designs, the waveguides and pinching locations are configured to deliver individual or group-specific data streams while controlling inter-user interference. In AirComp, by contrast, the superimposed signals of the participating devices are not interference but jointly constitute the desired aggregate.}
	
	The spatial reconfigurability of PASS makes it particularly attractive
	for AirComp-based systems \cite{PASSOTA}. Because AirComp aggregation relies heavily on favorable effective channels and precise signal alignment across devices, the ability to physically adjust radiating elements offers a powerful additional degree of freedom. This spatial flexibility allows the system to actively shape the received signal superposition, thereby substantially improving aggregation accuracy.
	Consequently, PASS is well suited for AirFL, where communication design
	directly impacts learning performance through aggregation distortion.
	Despite these advantages, the use of PASS in AirFL has received very
	limited attention. To the best of our knowledge, the closest existing work in the AirFL setting is
	\cite{PASSAirFL}, which considers an energy-efficiency objective under a
	single-waveguide PASS architecture. However, that work deviates from the
	widely studied learning-oriented maximum-participation AirFL formulation \cite{airflmag}. \textcolor{blue}{Hence, the design of PASS-based AirFL systems under this formulation remains largely unexplored.}
	
	Motivated by these observations, this paper investigates
	\emph{AirFL via pinching antenna systems}, referred to as
	AirPASS. We consider a wireless FL system in which the
	AP is equipped with a \textcolor{blue}{multiple-input multiple-output (MIMO)}
	PASS architecture consisting of multiple dielectric waveguides
	with movable pinching antennas. Selected devices simultaneously
	transmit their local model updates using AirComp, and the AP
	aggregates them via receive beamforming shaped by the positions
	of the pinching antennas across the waveguides. The main contributions of this paper are summarized as follows:
	\begin{itemize}
		\item We formulate a PASS-enabled AirFL design problem under the learning-oriented objective of maximizing the number of selected devices subject to an aggregation \textcolor{blue}{mean-square error (MSE)} constraint. The resulting formulation couples device selection, receive beamforming, and pinching-antenna placement in a multi-waveguide PASS receiver.
		
		\item For \textcolor{blue}{a} fixed PASS configuration, we reformulate the joint device-selection and receive-beamforming problem as a maximum-cardinality quadratic feasibility problem on the complex unit sphere. Based on this reformulation, we develop a homotopy-Riemannian margin-consolidation (HRMC) algorithm that combines smooth cardinality approximation, homotopy-reweighted Riemannian optimization, and active-set feasibility-margin consolidation. Unlike greedy scheduling methods and SDR-DC approaches that rely on matrix lifting, the proposed method performs all device-selection and receive-beamforming updates directly in the original beamforming space.
		
		\item For fixed selected devices and beamformer, we develop a homotopy-assisted geometry optimization (HAGO) method for PASS configuration. The proposed approach optimizes a smooth worst-device objective and employs a feasible reparameterization that enforces antenna ordering and minimum spacing by construction, enabling efficient joint continuous optimization of all pinching locations.
		
	\item We integrate the proposed HRMC and HAGO \textcolor{blue}{blocks} into the AirPASS alternating optimization framework\textcolor{blue}{, whose incumbent-retention rule guarantees feasible iterates, a nondecreasing number of selected devices, and finite termination. Experiments on MNIST and CIFAR-10 show that, with the same number of radio-frequency (RF) chains, AirPASS substantially increases device participation and improves learning performance over co-located MIMO and fixed-position PASS receivers, mainly owing to the geometry optimization, and attains a learning accuracy comparable to that of a co-located array with one RF chain per pinching antenna. Moreover, AirPASS matches the participation of SDR-DC to within $1$~dB in signal-to-noise ratio (SNR), with comparable accuracy, at about $2\%$ of its runtime for $50$ devices, and it clearly outperforms matching-pursuit scheduling.}
\end{itemize}
	
	\textcolor{blue}{The remainder of this paper is organized as follows. Section~II presents the system model, including the federated averaging (FedAvg) procedure, the MIMO PASS architecture, the uplink channel model, and the AirComp-based model aggregation. Section~III formulates the joint design problem and motivates the decomposition of its variables. Section~IV develops the HRMC algorithm for device selection and receive beamforming under a fixed PASS configuration. Section~V develops the HAGO algorithm for PASS configuration, integrates the two blocks into the AirPASS framework, and analyzes its monotonicity, finite termination, complexity, and operating timescale. Section~VI presents the experimental results, and Section~VII concludes the paper.}

\vspace{-10pt}

\section{System Model}

Consider a wireless FL system consisting of an AP and $K$ single-antenna edge devices indexed by the set $\mathcal{K}=\{1,\ldots,K\}$, as shown in Fig. 1. Device $k\in\mathcal{K}$ possesses a local dataset $\mathcal{D}_k$ with cardinality $|\mathcal{D}_k|$. The objective of FL is to collaboratively train a global model parameterized by $\mathbf{w}$ by minimizing the empirical risk
\begin{equation}\label{eq:global_risk}
	F(\mathbf{w})=\sum_{k=1}^{K}\frac{|\mathcal{D}_k|}{D}F_k(\mathbf{w}),
\end{equation}
where $F_k(\mathbf{w})$ denotes the local loss function of device $k$, and $D=\sum_{k=1}^{K}|\mathcal{D}_k|$.

Training proceeds over communication rounds following the FedAvg protocol \cite{McMahan}. At each round, the AP broadcasts the current global model \textcolor{blue}{$\mathbf w$} to all devices. Each device \textcolor{blue}{$k$ initializes its local model with $\mathbf w$, performs $J$ local mini-batch gradient steps on $F_k(\cdot)$ using $\mathcal D_k$, and computes its local update $\boldsymbol\Delta_k$ as the difference between the resulting local model and $\mathbf w$. For a set $\mathcal S$ of participating devices, FedAvg updates the global model with the data-weighted average of their local updates,
\[
\mathbf w\leftarrow\mathbf w+\sum_{k\in\mathcal S}\frac{|\mathcal D_k|}{D_{\mathcal S}}\boldsymbol\Delta_k,
\qquad
D_{\mathcal S}\triangleq\sum_{k\in\mathcal S}|\mathcal D_k|,
\]
which uses the weights of \eqref{eq:global_risk} when $\mathcal S=\mathcal K$. In the considered system, the devices in $\mathcal S$ transmit their updates simultaneously, and the weighted sum in this average is computed by AirComp, as described in Section~II-C.}

\begin{figure}[!t]
	\hspace{-20pt}
	\includegraphics[width=1.2\linewidth]{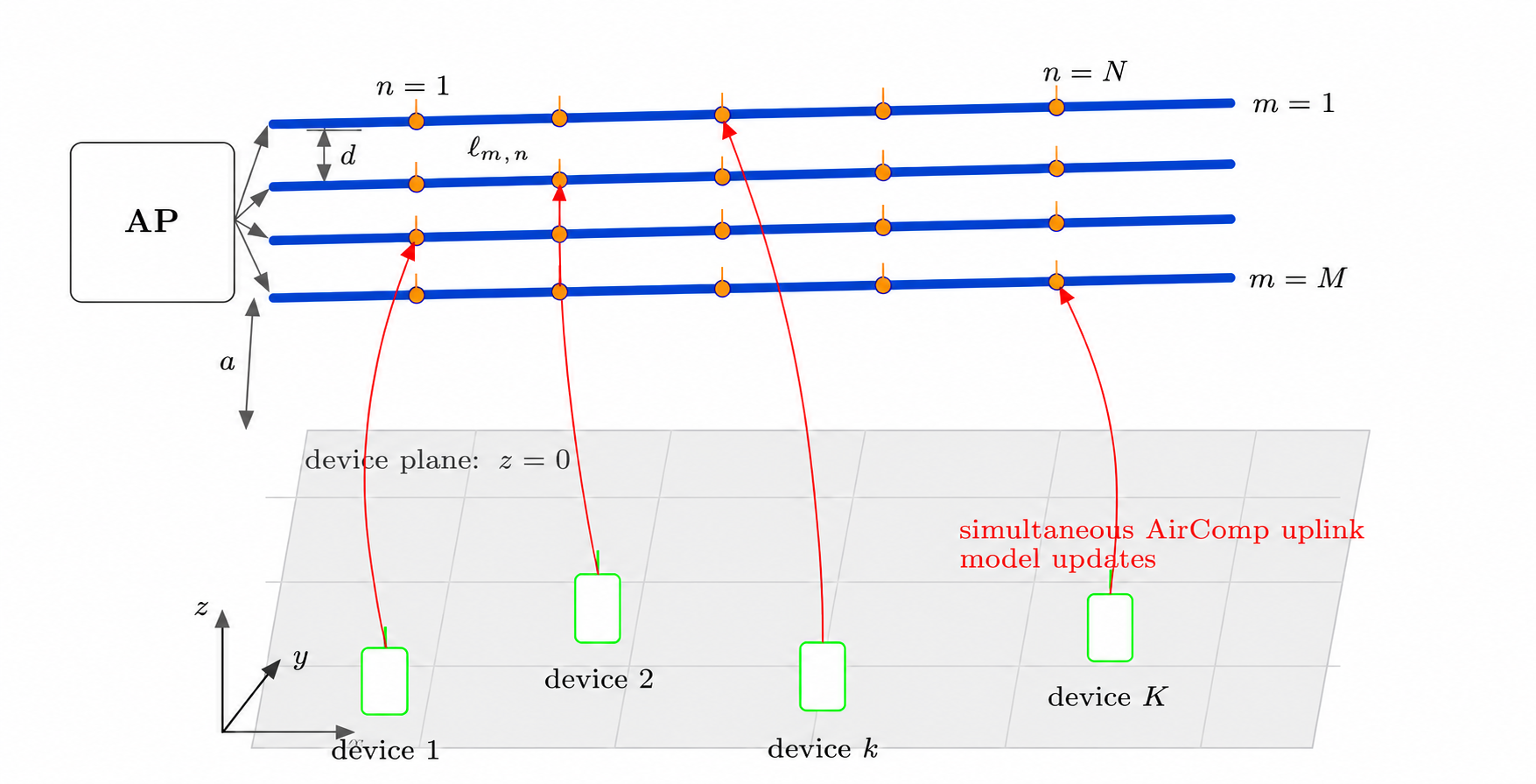}
	\caption{System model of the AirPASS-enabled wireless FL setup.}
	\label{fig:airpass_system_model}
	\vspace{-5pt}
\end{figure}
\vspace{-10pt}
\subsection{MIMO PASS Architecture}

The AP is equipped with a \textcolor{blue}{MIMO PASS} consisting of $M$ dielectric waveguides. Each waveguide is connected to one RF chain and contains $N$ movable pinching antennas. Accordingly, the AP provides an effective $M$-dimensional receive interface.

The $m$th waveguide is placed parallel to the $x$-axis at lateral coordinate \textcolor{blue}{$\bar y_m$} and height $a$\textcolor{blue}{, where $\bar y_1<\cdots<\bar y_M$}. Let $\ell_{m,n}$ denote the position of the $n$th pinching antenna on waveguide $m$ measured along the waveguide axis. The three-dimensional coordinate of the $n$th pinching antenna on waveguide $m$ is given by
\begin{equation}
	\mathbf{v}_{m,n}=[\ell_{m,n},{\color{blue}\bar y_m},a]^T,
\end{equation}
where $m\in\{1,\ldots,M\}$ and $n\in\{1,\ldots,N\}$.

The coordinate of device $k$ is denoted by
\begin{equation}
	\mathbf{u}_k=[x_k,y_k,0]^T.
\end{equation}

Define the pinching position vector on waveguide $m$ as
\begin{equation}
	\boldsymbol{\ell}_m=[\ell_{m,1},\ldots,\ell_{m,N}]^T,
\end{equation}
and collect all pinching locations as
\begin{equation}
	\mathbf{L}=[\boldsymbol{\ell}_1^T,\ldots,\boldsymbol{\ell}_M^T]^T.
\end{equation}

The pinching positions must satisfy the physical constraints
\begin{align}
	0 \le \ell_{m,1}, \quad &\forall m,\\
	\ell_{m,n+1}-\ell_{m,n}\ge \Delta \ell, \quad &\forall m,\; n=1,\ldots,N-1,\\
	\ell_{m,N}\le L_m, \quad &\forall m,
\end{align}
where $L_m$ denotes the length of waveguide $m$, and $\Delta \ell$ denotes the minimum allowable spacing between adjacent pinching antennas on the same waveguide, \textcolor{blue}{imposed to avoid} mutual coupling.

\vspace{-10pt}
\subsection{Uplink Channel Model}

The propagation distance between device $k$ and the $n$th pinching antenna on waveguide $m$ is
\begin{align}
	&	D_{k,m}(\ell_{m,n})=
	\|\mathbf{u}_k-\mathbf{v}_{m,n}\|
	= \nonumber\\&
	\sqrt{(x_k-\ell_{m,n})^2+(y_k-{\color{blue}\bar y_m})^2+a^2}.
\end{align}

Following the MIMO PASS uplink channel model \cite{MIMOPASS} \textcolor{blue}{and accounting for in-waveguide attenuation}, the effective channel coefficient between device $k$ and waveguide $m$ is
\begin{equation}\label{eq:channel}
	g_{k,m}(\boldsymbol{\ell}_m)
	=
	\xi \gamma_k
	\sum_{n=1}^{N}
	\frac{
		{\color{blue}e^{-\zeta\ell_{m,n}}}\,
		e^{-j\kappa\left(D_{k,m}(\ell_{m,n})+i_{\mathrm{ref}}\ell_{m,n}\right)}
	}{
		\sqrt{N}D_{k,m}(\ell_{m,n})
	},
\end{equation}
where $\xi$ is a constant related to the free-space propagation attenuation, $\gamma_k\in\mathbb{R}_+$ denotes the large-scale channel gain coefficient of device $k$, $\kappa=2\pi/\lambda$ is the wave number, $\lambda$ is the carrier wavelength, $i_{\mathrm{ref}}$ denotes the refractive index of the dielectric waveguide\textcolor{blue}{, and $\zeta\ge0$ is the in-waveguide attenuation coefficient, with the feed point of each waveguide located at $\ell=0$}.

\textcolor{blue}{The factor $e^{-\zeta\ell_{m,n}}$ captures the attenuation between the
$n$th pinching antenna and the feed point of waveguide $m$
\cite{PASSAttenuation,PASSAttenuationBlockage}. For typical dielectric
waveguides, the attenuation is about $0.08$~dB/m, i.e.,
$\zeta\approx0.0092$~m$^{-1}$
\cite{PASSAttenuation,PASSAttenuationBlockage}, so that the accumulated power
loss does not exceed $3.2$~dB over a $40$-m waveguide. These works also
quantify its effect on pinching-antenna design: the rate loss caused by
optimizing the antenna placement without accounting for in-waveguide
attenuation is negligible under line-of-sight propagation unless the service
area is excessively large \cite{PASSAttenuation}, and it remains negligible
under typical line-of-sight blockage even for large areas
\cite{PASSAttenuationBlockage}. Accordingly, the lossless case $\zeta=0$,
which recovers the model of \cite{MIMOPASS}, is used as the baseline in
Section~VI, and the effect of $\zeta=0.0092$~m$^{-1}$ is evaluated in
Section~\ref{sec:exp_ablation}. The proposed design, however, does not rely on this assumption:
the algorithms in Sections~IV and~V depend on the PASS geometry only through
$\mathbf g_k(\mathbf L)$ and its derivatives, and therefore apply to any
$\zeta\ge0$.}

The effective channel vector between device $k$ and the AP is defined as
\begin{equation}
	\mathbf{g}_k(\mathbf{L})
	=
	[g_{k,1}(\boldsymbol{\ell}_1),\ldots,g_{k,M}(\boldsymbol{\ell}_M)]^T
	\in \mathbb{C}^{M\times1}.
\end{equation}

Stacking all device channels yields the channel matrix
\begin{equation}
	\mathbf{G}(\mathbf{L})
	=
	[\mathbf{g}_1(\mathbf{L}),\ldots,\mathbf{g}_K(\mathbf{L})]
	\in \mathbb{C}^{M\times K}.
\end{equation}

\vspace{-10pt}
\subsection{AirComp-Based Model Aggregation}

At each communication round, only a subset of devices participates in the uplink transmission. Let $\mathcal{S}\subseteq\mathcal{K}$ denote the selected device set.

Each selected device $k\in\mathcal{S}$ transmits a scaled version of its normalized local update $s_k$\footnote{Here, $s_k$ denotes a scalar entry of the \textcolor{blue}{local update $\boldsymbol\Delta_k$ normalized with a mean and a standard deviation $\nu$ common to all selected devices; the AP inverts this normalization after aggregation}. As \textcolor{blue}{is} standard in AirComp-based FL, the same scalar formulation applies element-wise across model coordinates, and the resulting MSE expression characterizes the distortion of each aggregated entry.} as $\tilde{x}_k=b_k s_k$,
where $b_k$ is the transmit scaling coefficient subject to the power constraint $|b_k|^2\le P_0$.

The received signal at the AP is
\begin{equation}
	\mathbf{r}
	=
	\sum_{k\in\mathcal{S}}\mathbf{g}_k(\mathbf{L})b_k s_k
	+
	\mathbf{n},
\end{equation}
where $\mathbf{n}\sim\mathcal{CN}(\mathbf{0},\sigma^2\mathbf{I}_M)$ denotes the additive white Gaussian noise vector.

To recover the desired \textcolor{blue}{aggregate}, the AP applies a receive beamforming vector $\mathbf{m}\in\mathbb{C}^{M}$ and a denoising factor $\eta>0$, yielding
\begin{equation}
	\hat{s}=
	\frac{1}{\sqrt{\eta}}\mathbf{m}^H\mathbf{r}.
\end{equation}

In FL, the desired aggregation function is the weighted sum
\begin{equation}
	s=
	\sum_{k\in\mathcal{S}}\phi_k s_k,
\end{equation}
where $\phi_k={|\mathcal{D}_k|}$.

The AirComp aggregation distortion is measured by the mean-square error (MSE)
\begin{equation}
	\mathrm{MSE}=
	\mathbb{E}\left[|\hat{s}-s|^2\right].
\end{equation}

Assuming $\mathbb{E}[s_k]=0$, $\mathbb{E}[|s_k|^2]=1$, and \textcolor{blue}{uncorrelated $s_k$}, the MSE can be expressed as
\begin{equation}\label{eq:mse_expanded}
	\mathrm{MSE}
	=
	\sum_{k\in\mathcal{S}}
	\left|
	\frac{1}{\sqrt{\eta}}\mathbf{m}^H\mathbf{g}_k(\mathbf{L})b_k-\phi_k
	\right|^2
	+
	\frac{\sigma^2\|\mathbf{m}\|^2}{\eta}.
\end{equation}

\textcolor{blue}{Following the FedAvg update in Section~II, the AP divides
the recovered sum by $D_{\mathcal S}$, so that $\hat s/D_{\mathcal S}$
estimates the weighted average
$s/D_{\mathcal S}$ of the normalized entries, which the inverse normalization maps to the corresponding entry of the FedAvg update; ideal FedAvg corresponds to
$\mathcal S=\mathcal K$ and $\hat s=s$. The aggregation error of this estimate
is
\[
\mathrm{MSE}_{\mathrm{FedAvg}}
\triangleq
\mathbb{E}\!\left[\left|\frac{\hat s}{D_{\mathcal S}}-\frac{s}{D_{\mathcal S}}\right|^2\right]
=
\frac{\mathrm{MSE}}{D_{\mathcal S}^2}.
\]
Hence, the aggregation-MSE constraint $\mathrm{MSE}\le\varepsilon$ imposed in
Section~III, which is defined on the weighted sum $s$ as in \cite{DCSDR}, is
equivalent to $\mathrm{MSE}_{\mathrm{FedAvg}}\le\varepsilon/D_{\mathcal S}^2$.
Since the inverse normalization scales this error by $\nu^2$, the constraint bounds the error of the FedAvg update by a tolerance that
decreases as the selected devices contribute more data. This choice
keeps the per-device feasibility conditions
derived in Section~III independent of the selected set, whereas a fixed
tolerance $\bar\varepsilon$ on $\mathrm{MSE}_{\mathrm{FedAvg}}$ corresponds to
the set-dependent threshold $\varepsilon=\bar\varepsilon D_{\mathcal S}^2$,
which couples these conditions through $D_{\mathcal S}$. Moreover, replacing
$(\phi_k,\varepsilon,\eta)$ by $(c\phi_k,c^2\varepsilon,\eta/c^2)$ for any
$c>0$ scales the MSE in \eqref{eq:mse_expanded} by $c^2$ without changing
$\{b_k\}$, so the design depends on the dataset sizes only relative to
$\sqrt\varepsilon$: measuring $|\mathcal D_k|$ in any unit, e.g., the mean local
dataset size $\bar D=D/K$ used in Section~VI, is equivalent to expressing
$\varepsilon$ in the square of that unit.}

\section{Optimization Problem Formulation}

The objective of the proposed PASS-enabled AirFL system is to maximize the number of selected devices while ensuring that the aggregation MSE remains below a prescribed threshold. This formulation is \textcolor{blue}{motivated by convergence analyses of} AirFL systems \cite{airflmag} and has been widely adopted in the literature \cite{DCSDR,BeamformingVectorDesignFL,DeviceSchedulingOTAFL, IRSFL, isacfl}. At the communication-design level, this criterion requires channel information and aggregation weights, but does not require instantaneous gradients, losses, or class-distribution information. It therefore provides a practical model-agnostic scheduling surrogate for AirFL, while the final learning performance can still depend on the data partition and on the selected devices.

To this end, the AP jointly designs the device selection set $\mathcal{S}$, the transmit scaling coefficients $\{b_k\}$, the receive beamforming vector $\mathbf{m}$, the denoising factor $\eta$, and the pinching locations $\mathbf{L}$. The joint design problem is formulated as
\begin{subequations}\label{problem_main}
	\begin{align}
		\max_{\mathcal{S},\{b_k\},\mathbf{m},\eta,\mathbf{L}}
		\quad & |\mathcal{S}| \\
		\text{s.t.}\quad
		& \mathrm{MSE}\le \varepsilon, \\
		& |b_k|^2\le P_0,\quad \forall k\in\mathcal{S}, \\
		& 0 \le \ell_{m,1}, \quad \forall m, \\
		& \ell_{m,n+1}-\ell_{m,n}\ge \Delta \ell,\quad
		\forall m,\; n=1,\ldots,N-1, \\
		& \ell_{m,N}\le L_m,\quad \forall m.
	\end{align}
\end{subequations}

\textcolor{blue}{As in \cite{DCSDR}, we adopt zero-forcing transmit scaling, which, for any $k\in\mathcal{S}$ with $\mathbf{m}^H\mathbf{g}_k(\mathbf{L})\neq 0$, reads}
\begin{equation}
	b_k
	=
	\sqrt{\eta}\phi_k
	\frac{(\mathbf{m}^H\mathbf{g}_k(\mathbf{L}))^H}
	{|\mathbf{m}^H\mathbf{g}_k(\mathbf{L})|^2},
	\qquad k\in\mathcal{S}.
\end{equation}
\textcolor{blue}{This scaling} sets each effective noiseless aggregation coefficient to $\phi_k$, \textcolor{blue}{which eliminates the deterministic alignment error, i.e., the first summation term in \eqref{eq:mse_expanded}, so that $\hat s$ is an unbiased estimate of $s$ and the FedAvg weights are not distorted}. The remaining distortion is therefore purely noise-induced and decreases monotonically with $\eta$. Hence, \textcolor{blue}{under this design,} the optimal denoising factor is the largest one that satisfies the per-device power constraints.

Substituting the above expression into the transmit power constraint yields \cite{DCSDR}
\begin{equation}
	\eta
	\le
	\frac{P_0|\mathbf{m}^H\mathbf{g}_k(\mathbf{L})|^2}{\phi_k^2},
	\qquad \forall k\in\mathcal{S}.
\end{equation}

To satisfy all power constraints simultaneously, the denoising factor $\eta$ must satisfy \cite{DCSDR}
\begin{equation}\label{eq:eta_min}
	\eta
	\le
	\min_{k\in\mathcal{S}}
	\frac{P_0|\mathbf{m}^H\mathbf{g}_k(\mathbf{L})|^2}{\phi_k^2}.
\end{equation}

\textcolor{blue}{With $\eta$ equal to this bound, the minimum MSE under zero-forcing scaling becomes}
\begin{equation}\label{eq:mse_min}
	\mathrm{MSE}
	=
	\frac{\sigma^2}{P_0}
	\max_{k\in\mathcal{S}}
	\frac{\phi_k^2\|\mathbf{m}\|^2}
	{|\mathbf{m}^H\mathbf{g}_k(\mathbf{L})|^2}.
\end{equation}

Therefore, the aggregation-MSE constraint $\mathrm{MSE}\le\varepsilon$ is equivalent to
\begin{equation}
	\max_{k\in\mathcal{S}}
	\frac{\phi_k^2\|\mathbf{m}\|^2}
	{|\mathbf{m}^H\mathbf{g}_k(\mathbf{L})|^2}
	\le
	\frac{P_0\varepsilon}{\sigma^2}.
\end{equation}

Define
\begin{equation}
	\Gamma=\frac{P_0\varepsilon}{\sigma^2}.
\end{equation}

Then\textcolor{blue}{, under zero-forcing scaling,} the joint device selection, receive beamforming, and PASS configuration problem can be reformulated as
\begin{subequations}\label{problem_reformulated}
	\begin{align}
		\max_{\mathcal{S},\mathbf{m},\mathbf{L}}
		\quad & |\mathcal{S}| \\
		\text{s.t.}\quad
		&
		\frac{\phi_k^2\|\mathbf{m}\|^2}
		{|\mathbf{m}^H\mathbf{g}_k(\mathbf{L})|^2}
		\le
		\Gamma,
		\quad \forall k\in\mathcal{S}, \\
		&
		0 \le \ell_{m,1}, \quad \forall m, \\
		&
		\ell_{m,n+1}-\ell_{m,n}\ge \Delta \ell,
		\quad \forall m,\; n=1,\ldots,N-1, \\
		&
		\ell_{m,N}\le L_m,\quad \forall m.
	\end{align}
\end{subequations}

Problem \eqref{problem_reformulated} is a nonconvex and combinatorial optimization problem. \textcolor{blue}{The structurally different roles of its variables determine how they are grouped in the proposed AirPASS framework. Device selection cannot be meaningfully separated from receive beamforming: for fixed $(\mathbf m,\mathbf L)$, the largest feasible selection is simply the set of devices that satisfy their constraints in \eqref{problem_reformulated}, so that the joint selection--beamforming problem reduces to a cardinality problem in $\mathbf m$ alone. Alternating instead between thresholding for fixed $\mathbf m$ and beamforming for fixed $\mathcal S$ would optimize the beamformer only for the current set and admit additional devices only incidentally. The PASS geometry, in contrast, enters only through $\mathbf g_k(\mathbf L)$, in a highly nonconvex manner owing to the wavelength-scale phase terms in \eqref{eq:channel}, whereas $\mathbf m$ enters through quadratic forms. For fixed $(\mathcal S,\mathbf m)$, the natural geometry objective is the worst normalized effective channel of the selected devices, which determines whether they remain feasible. AirPASS therefore alternates between a selection--beamforming block for fixed $\mathbf L$ in Section~IV and a geometry block for fixed $(\mathcal S,\mathbf m)$ in Section~V. Compared with a fully joint update of $(\mathbf m,\mathbf L)$, this partition gives each block a structured problem (a sphere-constrained quadratic feasibility problem and a max--min problem over admissible configurations, respectively), enables the monotone scheme of Section~V-K, and allows the selected set and beamformer to be updated without moving the pinching antennas (Section~V-M).}

	\section{Device Selection and Receive Beamforming for Fixed PASS Configuration}

In this section, we study the joint device-selection and receive-beamforming
subproblem for a \emph{fixed} PASS configuration $\mathbf L$. For given
$\mathbf L$, the effective uplink channel vectors
$\{\mathbf g_k(\mathbf L)\}_{k\in\mathcal K}$ are known. The problem is
\begin{subequations}\label{problem_fixedL}
	\begin{align}
		\max_{\mathcal S,\mathbf m}\quad & |\mathcal S| \\
		\text{s.t.}\quad &
		\frac{\phi_k^2\|\mathbf m\|^2}
		{|\mathbf m^H\mathbf g_k(\mathbf L)|^2}
		\le \Gamma,
		\qquad \forall k\in\mathcal S .
	\end{align}
\end{subequations}

Problem \eqref{problem_fixedL} aims to select as many devices as possible while
satisfying the aggregation-MSE requirement. Devices with larger aggregation
weights $\phi_k$ or weaker effective channels $\mathbf g_k(\mathbf L)$ are
harder to schedule, since they require larger effective receive gains. The primary challenge in solving \eqref{problem_fixedL} stems from its combinatorial objective coupled with nonconvex quadratic feasibility constraints. 
\vspace{-5pt}
\subsection{Problem Reformulation and Main Idea}

The feasibility condition in \eqref{problem_fixedL} is homogeneous in
$\mathbf m$. Hence, we impose
\begin{equation}
	\|\mathbf m\|^2 = 1
\end{equation}
without loss of optimality. Define
\begin{equation}
	\mathbf Q_k(\mathbf L)
	=
	\frac{\Gamma}{\phi_k^2}\mathbf g_k(\mathbf L)\mathbf g_k^H(\mathbf L),
	\qquad \mathbf Q_k(\mathbf L)\succeq 0 .
\end{equation}
Then device $k$ is feasible if
\begin{equation}
	\mathbf m^H\mathbf Q_k(\mathbf L)\mathbf m \ge 1 .
\end{equation}

Accordingly, \eqref{problem_fixedL} can be rewritten as
\begin{subequations}\label{problem_sphere}
	\begin{align}
		\max_{\mathbf m}\quad &
		\sum_{k=1}^{K}
		\mathds{1}\!\left(
		\mathbf m^H\mathbf Q_k(\mathbf L)\mathbf m \ge 1
		\right) \\
		\text{s.t.}\quad & \|\mathbf m\|^2 = 1 .
	\end{align}
\end{subequations}

Problem \eqref{problem_sphere} is a maximum-cardinality quadratic feasibility
problem on the complex unit sphere. This reformulation is a central step of the
proposed method: it transforms the original mixed discrete--continuous problem
into a sphere-constrained cardinality formulation, making the beamformer
geometry explicit while isolating the combinatorial difficulty in the counting
objective.

Based on \eqref{problem_sphere}, we develop a
\emph{homotopy-Riemannian margin-consolidation (HRMC)} algorithm for device
selection and receive beamforming under \textcolor{blue}{a} fixed PASS configuration. The key idea is to
perform both the cardinality-oriented search and the final robustness-oriented
consolidation directly on the complex sphere. Thus, the fixed-configuration
block remains in the original beamforming space and avoids the lifted matrix
variables required by SDR-DC-type methods. 
\vspace{-5pt}
\subsection{Smooth Cardinality Approximation}

The main obstacle in \eqref{problem_sphere} is the discontinuous indicator
function. To handle it, we introduce a smooth surrogate that approximates the
number of violated constraints.

Define the feasibility score
\begin{equation}\label{eq:feas_score}
	z_k(\mathbf m;\mathbf L)
	=
	\mathbf m^H\mathbf Q_k(\mathbf L)\mathbf m ,
\end{equation}
and the one-sided violation measure
\begin{equation}
	v_k(\mathbf m)=\max\{1-z_k(\mathbf m;\mathbf L),0\}.
\end{equation}
Since the hinge operator is not differentiable, we replace it by the softplus
approximation
\begin{equation}\label{eq:softplus}
	\tilde v_k(\mathbf m)
	=
	\frac{1}{\beta}\log\!\left(
	1+\exp\!\big(\beta(1-z_k(\mathbf m;\mathbf L))\big)
	\right),
\end{equation}
where $\beta>0$ controls the sharpness. To mimic the counting behavior of the
cardinality objective, we further introduce the saturating penalty
\begin{equation}\label{eq:sat_penalty}
	\rho_\mu(t)=1-\exp(-t/\mu),
\end{equation}
where $\mu>0$ is a smoothing parameter.

This leads to the surrogate problem
\begin{subequations}\label{problem_smooth}
	\begin{align}
		\min_{\mathbf m}\quad &
		F_{\beta,\mu}(\mathbf m)
		=
		\sum_{k=1}^{K} w_k\,\rho_\mu(\tilde v_k(\mathbf m)) \\
		\text{s.t.}\quad & \|\mathbf m\|^2=1,
	\end{align}
\end{subequations}
where $w_k>0$ is the reweighting coefficient associated with device $k$. \textcolor{blue}{This surrogate} replaces
the discontinuous counting objective by a differentiable function while
preserving the selection structure of the original problem: small violations,
large violations, and feasible devices are treated differently. The reweighting
scheme assigns larger emphasis to devices with smaller surrogate violations,
thereby biasing the optimization toward recovering marginal or nearly feasible
devices.
\vspace{-5pt}
\subsection{Homotopy-Reweighted Riemannian Optimization}

Problem \eqref{problem_smooth} is naturally defined on the complex sphere
\[
\mathcal M=\{\mathbf m\in\mathbb C^M:\|\mathbf m\|=1\}.
\]
We solve it directly on $\mathcal M$, which preserves the unit-norm constraint
exactly and keeps the dominant optimization in the original beamforming space.

Using the Wirtinger convention for complex-valued variables, the derivative of
$z_k(\mathbf m;\mathbf L)$ with respect to $\mathbf m^\ast$ is
\begin{equation}
	\nabla_{\mathbf m^\ast}z_k(\mathbf m;\mathbf L)
	=
	\mathbf Q_k(\mathbf L)\mathbf m .
\end{equation}
Hence, the Euclidean gradient of \eqref{problem_smooth} is
\begin{equation}
	\nabla_{\mathbf m^\ast}F_{\beta,\mu}
	=
	-\sum_{k=1}^{K}\alpha_k\,\mathbf Q_k(\mathbf L)\mathbf m,
\end{equation}
where
\[
\alpha_k
=
w_k\,\rho_\mu'(\tilde v_k)\,
\frac{1}{1+\exp\!\big(-\beta(1-z_k(\mathbf m;\mathbf L))\big)},
\]
and
\[
\rho_\mu'(t)=\frac{1}{\mu}\exp(-t/\mu).
\]
Projecting onto the tangent space gives the Riemannian gradient
\begin{equation}
	\mathrm{grad}\,F_{\beta,\mu}
	=
	\nabla_{\mathbf m^\ast}F_{\beta,\mu}
	-
	\mathbf m\Re\!\left\{
	\mathbf m^H\nabla_{\mathbf m^\ast}F_{\beta,\mu}
	\right\}.
\end{equation}

For initialization, we use the normalized dominant eigenvector of
$\sum_{k=1}^{K}\mathbf Q_k(\mathbf L)$. For fixed $(\beta,\mu)$ and weights
$\{w_k\}$, we solve \eqref{problem_smooth} by the standard Riemannian
conjugate-gradient \textcolor{blue}{(RCG)} method on $\mathcal M$ \cite{absil2008optimization}. At each
iteration, the method constructs a conjugate search direction
$\boldsymbol{\xi}$ in the tangent space of the current point. Given
$\boldsymbol{\xi}$ and a step size $\tau>0$, the next beamformer is obtained
by the normalized retraction
\begin{equation}
	\mathbf m^+
	=
	\mathcal R_{\mathbf m}(\tau\boldsymbol{\xi})
	=
	\frac{\mathbf m+\tau\boldsymbol{\xi}}
	{\|\mathbf m+\tau\boldsymbol{\xi}\|},
\end{equation}
where $\tau$ is chosen by backtracking line search.

A fixed surrogate pair $(\beta,\mu)$ generally does not provide a good balance
between numerical stability and approximation accuracy. We therefore use a
homotopy, or continuation, strategy: the algorithm solves a sequence of smooth
surrogate problems, each initialized from the previous solution, while gradually
sharpening the approximation to the original nonsmooth cardinality objective.
In HRMC, this is achieved by increasing the softplus sharpness and decreasing
the smoothing level of the saturating violation penalty\textcolor{blue}{, by factors $\kappa_\beta,\kappa_\mu>1$ until they reach $\beta_{\max}$ and $\mu_{\min}$}:
\begin{equation}\label{eq:homotopy_hrmc}
	\beta^{(t+1)}=\min\{\kappa_\beta\beta^{(t)},\beta_{\max}\},
	\mu^{(t+1)}=\max\{\mu^{(t)}/\kappa_\mu,\mu_{\min}\}.
\end{equation}
The weights are initialized as $w_k^{(0)}=1$ and updated as
\begin{equation}\label{eq_weight_update}
	w_k^{(t+1)}
	=
	\frac{1}{\tilde v_k(\mathbf m^{(t)})+\epsilon_w},
\end{equation}
where $\mathbf m^{(t)}$ denotes the beamformer obtained at homotopy step $t$
\textcolor{blue}{and $\epsilon_w>0$ is a regularizer}.
This rule implements the reweighting mechanism introduced in
\eqref{problem_smooth}.

After convergence of the homotopy iterations, let $\bar{\mathbf m}$ denote the
resulting beamformer and define
\[
\bar{\mathcal S}
=
\{k:\bar{\mathbf m}^H\mathbf Q_k(\mathbf L)\bar{\mathbf m}\ge 1\}.
\]

\subsection{Feasibility-Margin Consolidation}

The homotopy-reweighted stage identifies a beamforming direction that is
favorable for device admission. However, some feasible or nearly feasible
devices may remain close to the threshold. To improve the robustness of this
candidate set, we perform a feasibility-margin consolidation step on the same
complex sphere.

Let $\bar{\mathbf m}$ be the beamformer obtained from the homotopy stage, and
define
\[
z_k(\bar{\mathbf m};\mathbf L)
=
\bar{\mathbf m}^H\mathbf Q_k(\mathbf L)\bar{\mathbf m}.
\]
We form the active candidate set
\begin{equation}\label{eq:active_set}
	\mathcal A
	=
	\{k:z_k(\bar{\mathbf m};\mathbf L)\ge 1-\delta\},
\end{equation}
where $\delta\textcolor{blue}{\in(0,1)}$ is a margin parameter. The set $\mathcal A$ contains
devices that are already feasible under $\bar{\mathbf m}$, as well as devices
within a margin $\delta$ of feasibility; these are the devices targeted by the
local margin-consolidation step.

For $\mathcal A\neq\emptyset$, the weakest active feasibility score is
$\min_{k\in\mathcal A}\mathbf m^H\mathbf Q_k(\mathbf L)\mathbf m$. Since this
minimum is nondifferentiable, we replace it with the following smooth soft-min
approximation and maximize it:
\begin{subequations}\label{problem_margin}
	\begin{align}
		\max_{\mathbf m}\quad &
		F_{\mathrm{mc}}(\mathbf m)
		= -\tau_{\mathrm m}
		\log\!\left(
		\sum_{k\in\mathcal A}
		\exp\!\left(
		-\frac{\mathbf m^H\mathbf Q_k(\mathbf L)\mathbf m}{\tau_{\mathrm m}}
		\right)
		\right) \\
		\text{s.t.}\quad & \|\mathbf m\|^2=1,
	\end{align}
\end{subequations}
where $\tau_{\mathrm m}>0$ controls the accuracy of the soft-min approximation.
As $\tau_{\mathrm m}$ decreases, $F_{\mathrm{mc}}(\mathbf m)$ approaches
$\min_{k\in\mathcal A}\mathbf m^H\mathbf Q_k(\mathbf L)\mathbf m$.

Define
\begin{equation}
	\pi_k(\mathbf m)
	=
	\frac{
		\exp\!\left(
		-\mathbf m^H\mathbf Q_k(\mathbf L)\mathbf m/\tau_{\mathrm m}
		\right)}{
		\sum_{j\in\mathcal A}
		\exp\!\left(
		-\mathbf m^H\mathbf Q_j(\mathbf L)\mathbf m/\tau_{\mathrm m}
		\right)},
	\qquad k\in\mathcal A .
\end{equation}
The Euclidean gradient of the margin-consolidation objective is
\begin{equation}
	\nabla_{\mathbf m^\ast}F_{\mathrm{mc}}(\mathbf m)
	=
	\sum_{k\in\mathcal A}
	\pi_k(\mathbf m)\mathbf Q_k(\mathbf L)\mathbf m,
\end{equation}
and the corresponding Riemannian gradient is
\begin{equation}
	\mathrm{grad}\,F_{\mathrm{mc}}
	=
	\nabla_{\mathbf m^\ast}F_{\mathrm{mc}}(\mathbf m)
	-
	\mathbf m\Re\!\left\{
	\mathbf m^H\nabla_{\mathbf m^\ast}F_{\mathrm{mc}}(\mathbf m)
	\right\}.
\end{equation}
We solve \eqref{problem_margin} using the same \textcolor{blue}{RCG method} and normalized retraction as in the homotopy
stage, applied equivalently to the minimization of $-F_{\mathrm{mc}}$. If
$\mathcal A=\emptyset$, the consolidation step is skipped and
$\mathbf m^\star=\bar{\mathbf m}$. Otherwise, let $\mathbf m_{\mathrm{mc}}$
denote the beamformer obtained from \eqref{problem_margin}, and define
$\mathcal S_{\mathrm{mc}}=\{k:(\mathbf m_{\mathrm{mc}})^H
\mathbf Q_k(\mathbf L)\mathbf m_{\mathrm{mc}}\ge 1\}$. We set
$\mathbf m^\star=\mathbf m_{\mathrm{mc}}$ if
$|\mathcal S_{\mathrm{mc}}|\ge|\bar{\mathcal S}|$, and set
$\mathbf m^\star=\bar{\mathbf m}$ otherwise.

The final selected set is determined by the original feasibility condition:
\begin{equation}
\mathcal S^\star
=
\{k:(\mathbf m^\star)^H\mathbf Q_k(\mathbf L)\mathbf m^\star\ge 1\}.
\end{equation}
\vspace{-5pt}
\subsection{Proposed HRMC Algorithm}

Algorithm~\ref{alg:hybrid} summarizes the proposed HRMC algorithm for fixed
PASS configuration. The method consists of two coordinated stages:
homotopy-reweighted Riemannian optimization and active-set feasibility-margin
consolidation. The first stage performs the main cardinality-oriented
beamformer search, while the second improves the robustness of the active
candidate set before the final thresholding decision.

\begin{algorithm}[t]
\caption{HRMC Device Selection and Receive Beamforming Algorithm}
\label{alg:hybrid}
\begin{algorithmic}[1]
\State Initialize \textcolor{blue}{$\mathbf m$} as the normalized dominant eigenvector of
$\sum_{k=1}^{K}\mathbf Q_k(\mathbf L)$
\State Choose parameters $\beta,\mu,\epsilon_w,\kappa_\beta,\kappa_\mu,\beta_{\max},\mu_{\min},\tau_{\mathrm m},\delta$
\State Set $w_k^{(0)}=1,\;\forall k$

\Repeat
\State Solve \eqref{problem_smooth} via Riemannian conjugate gradient
\State Update the weights $w_k$
\State Update the homotopy parameters $\beta,\mu$
\Until{convergence}

\State Obtain $\bar{\mathbf m}$ and form
$\bar{\mathcal S}=\{k:\bar{\mathbf m}^H\mathbf Q_k(\mathbf L)\bar{\mathbf m}\ge 1\}$
\State Form the active candidate set
$\mathcal A=\{k:\bar{\mathbf m}^H\mathbf Q_k(\mathbf L)\bar{\mathbf m}\ge 1-\delta\}$

\If{$\mathcal A\neq\emptyset$}
\State Solve \eqref{problem_margin} over $\mathcal A$ via Riemannian conjugate gradient applied to minimizing $-F_{\mathrm{mc}}$
\State Let $\mathbf m_{\mathrm{mc}}$ denote the resulting beamformer and form
$\mathcal S_{\mathrm{mc}}=\{k:(\mathbf m_{\mathrm{mc}})^H\mathbf Q_k(\mathbf L)\mathbf m_{\mathrm{mc}}\ge 1\}$
\State Set $\mathbf m^\star=\mathbf m_{\mathrm{mc}}$ if
$|\mathcal S_{\mathrm{mc}}|\ge|\bar{\mathcal S}|$; otherwise set
$\mathbf m^\star=\bar{\mathbf m}$
\Else
\State $\mathbf m^\star=\bar{\mathbf m}$
\EndIf

\State Compute
$\mathcal S^\star=\{k:(\mathbf m^\star)^H\mathbf Q_k(\mathbf L)\mathbf m^\star\ge 1\}$

\State \textbf{Output:} $\mathbf m^\star,\mathcal S^\star$
\end{algorithmic}
\end{algorithm}

\subsection{Complexity Analysis}

The proposed HRMC block has two computational stages.

The first stage solves \eqref{problem_smooth} by homotopy-reweighted
Riemannian optimization. In each iteration, the dominant cost is evaluating
$z_k(\mathbf m;\mathbf L)$ and $\mathbf Q_k(\mathbf L)\mathbf m$ for all $k$.
Under a dense implementation, this gives $\mathcal O(KM^2)$ per iteration.
Since
\[
\mathbf Q_k(\mathbf L)
=
\frac{\Gamma}{\phi_k^2}\mathbf g_k(\mathbf L)\mathbf g_k^H(\mathbf L)
\]
is rank one, a structure-aware implementation reduces this to
$\mathcal O(KM)$.

The second stage solves the margin-consolidation problem
\eqref{problem_margin}. Its per-iteration complexity is
$\mathcal O(|\mathcal A|M^2)$ under a dense implementation and
$\mathcal O(|\mathcal A|M)$ when the rank-one structure is exploited. Since
$|\mathcal A|\le K$, the per-iteration cost of this stage is upper bounded by
that of the first stage.

Therefore, if the homotopy-reweighted stage and the margin-consolidation stage
require $I_1$ and $I_2$ Riemannian iterations, respectively, a conservative
overall bound is
\begin{equation}
\mathcal O\!\left(
I_1KM^2 + I_2|\mathcal A|M^2
\right),
\end{equation}
which yields
\begin{equation}
\mathcal O\!\left(
(I_1+I_2)KM^2
\right)
\end{equation}
using $|\mathcal A|\le K$. If the rank-one structure is exploited, the
corresponding bound becomes
\begin{equation}
\mathcal O\!\left(
I_1KM + I_2|\mathcal A|M
\right),
\end{equation}
and hence
\begin{equation}
\mathcal O\!\left(
(I_1+I_2)KM
\right).
\end{equation}
\vspace{-10pt}

\subsection{Discussion}
The fixed-configuration subproblem in \eqref{problem_fixedL} has the same essential
structure as the learning-oriented AirFL scheduling formulations studied in
prior work, namely, maximizing the number of selected devices under an
aggregation-MSE requirement. Existing approaches to this class of problems have
mainly followed two directions. One direction relies on lifted SDR-DC
formulations, where matrix optimization is the main computational tool
\cite{DCSDR}. Another direction relies on matching-pursuit-type scheduling,
which reduces complexity through sequential support updates
\cite{DeviceSchedulingOTAFL}.\footnote{\textcolor{blue}{These} prior methods
operate under fixed receiver architectures and do not incorporate PASS-related
design variables. In contrast, the present work embeds this subproblem within a
broader joint optimization that includes PASS configuration.}

\textcolor{blue}{A lightweight yet coupled HRMC block is especially
important} in the proposed alternating framework, where device selection and
receive beamforming are updated repeatedly as the PASS configuration changes. In this
setting, lifted high-complexity solvers can dominate the overall computational
cost, while purely sequential hard-selection rules may produce support updates
that are sensitive to intermediate PASS configurations.

The proposed HRMC method takes a different route. For a fixed PASS configuration,
the problem is first recast as a sphere-constrained maximum-cardinality
quadratic feasibility problem, so that the dominant search is carried out
directly in the receive-beamforming space. The subsequent margin-consolidation
stage remains on the same manifold and improves the weakest active feasibility
scores before the final device-selection decision. Thus, the method avoids the
lifted matrix variables used by SDR-DC approaches while retaining a coupled
continuous beamformer optimization that is absent from purely greedy scheduling.

\begin{table}[t]
	\centering
	\caption{Complexity comparison for fixed-configuration AirFL device-selection methods}
	\label{tab:complexity_comparison}
	\begin{tabular}{|c|c|}
		\hline
		\textbf{Method} & \textbf{Computational Complexity} \\
		\hline
		SDR-DC \cite{DCSDR} 
		& $\mathcal{O}\!\left(K(M^2+K)^3 + K M^6\right)$ \\
		\hline
		HRMC
		& $\mathcal{O}\!\left((I_1+I_2)K M^2\right)$ \\
		\hline
		Matching Pursuit \cite{DeviceSchedulingOTAFL} 
		& $\mathcal{O}\!\left(K^2 M^2\right)$ \\
		\hline
	\end{tabular}
	\vspace{-15pt}
\end{table}

The complexity results in Table~\ref{tab:complexity_comparison} highlight distinct scaling behaviors. 
The SDR-DC approach scales cubically in $(M^2+K)$ and includes an additional $M^6$ term, making it highly sensitive to the receive dimension. Matching pursuit scales quadratically in both $K$ and $M$, resulting in lower growth with system size but relying on sequential support decisions. The proposed HRMC method replaces the lifted SDR-DC cost by linear scaling in
$K$ and quadratic scaling in $M$ under the dense implementation reported in the
table, with a reduced $\mathcal O((I_1+I_2)KM)$ implementation possible when the
rank-one structure of $\mathbf Q_k(\mathbf L)$ is exploited. Consequently, HRMC
offers a lightweight manifold-based block for device selection and receive beamforming
within the \textcolor{blue}{AirPASS framework}.

\vspace{-5pt}

\section{PASS Antenna Configuration Optimization}

We now address the second block of the AirPASS framework, namely the
optimization of the \textcolor{blue}{pinching-antenna positions for a fixed device set
and receive beamformer}. This block is referred to as
\emph{homotopy-assisted geometry optimization (HAGO)}. Its role is to
reshape the effective uplink channels through antenna placement so as to
improve the aggregation feasibility margins of the selected devices.
\vspace{-5pt}
\subsection{PASS Configuration Optimization Problem}

From the aggregation feasibility condition derived earlier, we
have
\begin{equation}
	\frac{\phi_k^2\|\mathbf m\|^2}{|\mathbf m^H\mathbf g_k(\mathbf L)|^2}
	\le \Gamma ,\quad \forall k\in\mathcal S.
\end{equation}
Since the beamformer is normalized as $\|\mathbf m\|=1$, this condition
reduces to
\begin{equation}\label{eq:feas_pass}
	|\mathbf m^H\mathbf g_k(\mathbf L)|^2
	\ge
	\frac{\phi_k^2}{\Gamma},\quad \forall k\in\mathcal S.
\end{equation}

For fixed $(\mathcal S,\mathbf m)$, \textcolor{blue}{the objective of
\eqref{problem_reformulated} is constant, so} the PASS-configuration
subproblem \textcolor{blue}{amounts to} determining antenna positions
$\mathbf L$ that satisfy \textcolor{blue}{\eqref{eq:feas_pass}} for all selected
devices. \textcolor{blue}{These conditions hold jointly if and only if the
weakest normalized effective channel among the selected devices is at least
$1/\Gamma$, i.e., if and only if the worst of the feasibility scores in
\eqref{eq:feas_score} over the nonempty set $\mathcal S$ satisfies
\[
\psi(\mathcal S,\mathbf m,\mathbf L)
\triangleq
\min_{k\in\mathcal S}z_k(\mathbf m;\mathbf L)
=
\Gamma\min_{k\in\mathcal S}
\frac{|\mathbf m^H\mathbf g_k(\mathbf L)|^2}{\phi_k^2}
\ge 1.
\]
The same quantity determines the aggregation quality: the
denoising factor is common to all selected devices, so that
\eqref{eq:eta_min} limits it by the weakest normalized effective channel, and
the minimum MSE in \eqref{eq:mse_min} equals
$\varepsilon/\psi(\mathcal S,\mathbf m,\mathbf L)$. Maximizing the weakest
normalized effective channel over $\mathbf L$ therefore attains a feasible
configuration whenever one exists, minimizes the aggregation MSE of the
current selection, and maximizes its feasibility margin. This yields the
max--min subproblem}
\begin{subequations}\label{prob_pass}
	\begin{align}
		\max_{\mathbf L} \quad &
		\min_{k\in\mathcal S}
		\frac{|\mathbf m^H\mathbf g_k(\mathbf L)|^2}{\phi_k^2} \\
		\text{s.t.}\quad
		&0\le \ell_{m,1} \le \ell_{m,2}\le \cdots \le \ell_{m,N}\le L_m,\quad \forall m, \\
		&\ell_{m,n+1}-\ell_{m,n}\ge \Delta \ell,\quad
		\forall m,\; n=1,\ldots,N-1 .
	\end{align}
\end{subequations}


\textcolor{blue}{Problem} \eqref{prob_pass} is highly nonconvex because the effective
channel coefficients depend on antenna positions through nonlinear distance
and phase terms. In addition, the minimum operator in the objective is
nondifferentiable. The main challenge is therefore to optimize a
worst-device-driven configuration criterion under coupled ordering and
minimum-spacing constraints on all pinching locations.
\vspace{-5pt}
\subsection{Smooth Approximation of the Worst-Device Objective}

To enable efficient optimization, we replace the minimum operator in
\eqref{prob_pass} with a smooth log-sum-exp approximation. Define
\begin{equation}
	h_k(\mathbf L)=|\mathbf m^H\mathbf g_k(\mathbf L)|^2 .
\end{equation}
\textcolor{blue}{We then introduce} the smoothed objective
\begin{equation}\label{eq:lse_hago}
	f_\tau(\mathbf L)
	=
	-\tau
	\log
	\sum_{k\in\mathcal S}
	\exp
	\left(
	-\frac{h_k(\mathbf L)}{\tau\phi_k^2}
	\right),
\end{equation}
where $\tau>0$ is the smoothing parameter. It follows that
\begin{equation}
	\lim_{\tau\rightarrow 0}
	f_\tau(\mathbf L)
	=
	\min_{k\in\mathcal S}
	\frac{h_k(\mathbf L)}{\phi_k^2}.
\end{equation}
Hence, maximizing $f_\tau(\mathbf L)$ provides a smooth approximation of the
original worst-device objective.

For large $\tau$, the surrogate is smoother and easier to optimize, while
for small $\tau$ it becomes a tighter approximation of the minimum operator.
This motivates \textcolor{blue}{the} continuation strategy described next.

\subsection{Homotopy Strategy for the Smoothing Parameter}

Following the continuation principle introduced in the HRMC block, HAGO also
uses a homotopy schedule for its smoothing parameter. The algorithm
starts from a relatively large \textcolor{blue}{value $\tau_0$} and gradually decreases it \textcolor{blue}{by a
factor $\rho\in(0,1)$ to $\tau_{\min}$}, so that it first explores the PASS configuration under a smooth
objective landscape and then progressively focuses on the weakest effective
channel.

Specifically, the smoothing parameter is updated as
\begin{equation}\label{eq:homotopy_hago}
	\tau^{(t+1)}=\max\{\rho\,\tau^{(t)},\tau_{\min}\}.
\end{equation}
\vspace{-5pt}
\subsection{Gradient of the Smoothed Objective}

Define
\begin{equation}
	u_k(\mathbf L)=\mathbf m^H\mathbf g_k(\mathbf L),
\end{equation}
so that $h_k(\mathbf L)=|u_k(\mathbf L)|^2$. The derivative of $h_k$ with
respect to $\ell_{m,n}$ is
\begin{equation}
	\frac{\partial h_k}{\partial\ell_{m,n}}
	=
	2\Re
	\left\{
	u_k^\ast
	\mathbf m^H
	\frac{\partial\mathbf g_k(\mathbf L)}{\partial\ell_{m,n}}
	\right\}.
\end{equation}

Applying the chain rule to the log-sum-exp surrogate yields
\begin{equation}
	\frac{\partial f_\tau}{\partial\ell_{m,n}}
	=
	2
	\sum_{k\in\mathcal S}
	\widetilde w_k
	\Re
	\left\{
	u_k^\ast
	\mathbf m^H
	\frac{\partial\mathbf g_k(\mathbf L)}{\partial\ell_{m,n}}
	\right\},
\end{equation}
where
\begin{equation}
	\widetilde w_k
	=
	\frac{\exp(-h_k(\mathbf L)/(\tau\phi_k^2))}
	{\sum_{j\in\mathcal S}\exp(-h_j(\mathbf L)/(\tau\phi_j^2))}
	\cdot\frac{1}{\phi_k^2}.
\end{equation}
These weights emphasize devices with weaker normalized effective channels and
therefore steer the configuration update toward improving the bottleneck device.
\vspace{-5pt}
\subsection{Derivative of the PASS Channel}

The effective channel coefficient between device $k$ and waveguide $m$ is
\begin{equation}
	g_{k,m}(\boldsymbol{\ell}_m)
	=
	\sum_{n=1}^{N} g_{k,m,n},
\end{equation}
where
\begin{equation}
	g_{k,m,n}
	=
	\xi\gamma_k
	\frac{
		{\color{blue}e^{-\zeta\ell_{m,n}}}\,
		e^{-j\kappa(D_{k,m}(\ell_{m,n})+i_{\mathrm{ref}}\ell_{m,n})}
	}{
		\sqrt N\, D_{k,m}(\ell_{m,n})
	}.
\end{equation}
The propagation distance is
\begin{equation}
	D_{k,m}(\ell_{m,n})
	=
	\sqrt{(x_k-\ell_{m,n})^2+(y_k-{\color{blue}\bar y_m})^2+a^2}.
\end{equation}
Its derivative is
\begin{equation}
	\frac{\partial D_{k,m}}{\partial\ell_{m,n}}
	=
	\frac{\ell_{m,n}-x_k}{D_{k,m}(\ell_{m,n})}.
\end{equation}

Therefore,
\begin{align}
	&\frac{\partial g_{k,m,n}}{\partial\ell_{m,n}}
	=\nonumber\\&
	g_{k,m,n}
	\Bigg[
	-\frac{1}{D_{k,m}(\ell_{m,n})}
	\frac{\partial D_{k,m}}{\partial\ell_{m,n}}
	{\color{blue}-\zeta}
	-j\kappa
	\left(
	\frac{\partial D_{k,m}}{\partial\ell_{m,n}}
	+i_{\mathrm{ref}}
	\right)
	\Bigg].
\end{align}

Since only the $m$th waveguide component of $\mathbf g_k(\mathbf L)$ depends on
$\ell_{m,n}$, we have
\begin{equation}
	\frac{\partial \mathbf g_k(\mathbf L)}{\partial\ell_{m,n}}
	=
	\mathbf e_m
	\frac{\partial g_{k,m,n}}{\partial\ell_{m,n}},
\end{equation}
where $\mathbf e_m$ is the $m$th canonical basis vector in $\mathbb C^M$.

\subsection{Feasible Reparameterization of Antenna Positions}

Direct optimization over the antenna positions $\{\ell_{m,n}\}$ is inconvenient
because of the ordering and minimum-spacing constraints in \eqref{prob_pass}.
To address this issue, we introduce a reparameterization that guarantees
feasibility by construction. \textcolor{blue}{It}
converts the constrained configuration design into a smooth unconstrained problem
while preserving the joint coupling among all pinching locations.

For each waveguide $m$, define $N+1$ unconstrained real variables
$\{u_{m,i}\}_{i=0}^{N}$. We first map them to positive quantities through the
softplus function
\begin{equation}\label{eq:softplus_map}
	r_{m,i}=\log(1+e^{u_{m,i}}), \qquad i=0,1,\ldots,N .
\end{equation}
Let
\begin{equation}
	T_m=L_m-(N-1)\Delta \ell ,
\end{equation}
which is the remaining allocable length after accounting for the mandatory
minimum spacings. A feasible configuration requires $T_m\ge 0$.

Next, define normalized slack fractions
\begin{equation}\label{eq:slack_fraction}
	q_{m,i}=\frac{r_{m,i}}{\sum_{j=0}^{N} r_{m,j}},\qquad i=0,1,\ldots,N,
\end{equation}
so that $q_{m,i}>0$ and $\sum_{i=0}^{N} q_{m,i}=1$.

The antenna positions are then constructed as
\begin{equation}
	\ell_{m,n}
	=
	T_m\sum_{i=0}^{n-1} q_{m,i} + (n-1)\Delta \ell,
	\qquad n=1,\ldots,N .
\end{equation}

Under this mapping, the quantities $T_m q_{m,0}$ and $T_m q_{m,N}$ represent
the left and right endpoint slacks, respectively, while
$T_m q_{m,1},\ldots,T_m q_{m,N-1}$ represent the additional movable gaps
between adjacent antennas beyond the mandatory spacing $\Delta \ell$.
Therefore,
\begin{equation}
	0\le \ell_{m,1}\le \ell_{m,2}\le \cdots \le \ell_{m,N}\le L_m
\end{equation}
and
\begin{equation}
	\ell_{m,n+1}-\ell_{m,n}\ge \Delta \ell,\qquad n=1,\ldots,N-1,
\end{equation}
hold automatically for every waveguide. Since the softplus mapping gives
strictly positive slacks, this \textcolor{blue}{mapping parameterizes} the relative
interior of the feasible set; boundary configurations such as
$\ell_{m,1}=0$, $\ell_{m,N}=L_m$, or
$\ell_{m,n+1}-\ell_{m,n}=\Delta\ell$ can be approached arbitrarily closely
as limiting cases. Hence, \eqref{prob_pass} is converted into an unconstrained
optimization over the variables
$\mathbf U=\{u_{m,i}\}_{m=1,\ldots,M;\,i=0,\ldots,N}$.
\vspace{-5pt}
\subsection{Geometry Optimization in the Reparameterized Variables}

Using the above mapping, the smoothed objective becomes
$f_\tau(\mathbf L(\mathbf U))$, which is differentiable with respect to
$\mathbf U$. Therefore, the configuration update can be carried out by gradient
ascent in the reparameterized variables. By the chain rule,
\begin{equation}
	\frac{\partial f_\tau}{\partial \textcolor{blue}{u_{m,j}}}
	=
	\sum_{i=1}^{N}
	\frac{\partial f_\tau}{\partial \ell_{m,i}}
	\frac{\partial \ell_{m,i}}{\partial \textcolor{blue}{u_{m,j}}},
\end{equation}
where the derivatives $\partial f_\tau/\partial \ell_{m,i}$ are obtained
from \textcolor{blue}{Sections~V-D and~V-E}, while
$\partial \ell_{m,i}/\partial \textcolor{blue}{u_{m,j}}$ follow from the softplus and
normalization mapping. Since these terms are smooth, standard backtracking
line search can be used to determine the ascent step size.

The update is therefore
\begin{equation}\label{eq:hago_step}
	\mathbf U^{(t+1)}
	=
	\mathbf U^{(t)}+\alpha_t \nabla_{\mathbf U}
	f_{\tau^{(t)}}\big(\mathbf L(\mathbf U^{(t)})\big),
\end{equation}
where $\alpha_t>0$ \textcolor{blue}{is this step size}.
\vspace{-5pt}
\subsection{PASS Configuration Algorithm}

The resulting configuration optimization procedure is summarized in
Algorithm~\ref{alg_pass}. The algorithm alternates between computing the
gradient of the smoothed worst-device objective with respect to the
reparameterized variables and updating the smoothing parameter according to
the homotopy rule.

\begin{algorithm}[t]
	\caption{HAGO Geometry Algorithm}
	\label{alg_pass}
	\begin{algorithmic}[1]
		\State Initialize $\mathbf U^{(0)}$ \textcolor{blue}{such that $\mathbf L(\mathbf U^{(0)})$ is the current configuration}
	\State Choose $\tau_0,\rho,\tau_{\min}$ and initialize $\tau=\tau_0$
		\State Set iteration index $t=0$
		\Repeat
		\State Compute $\mathbf L(\mathbf U^{(t)})$
		\State Evaluate $f_{\tau}(\mathbf L(\mathbf U^{(t)}))$ and
		$\nabla_{\mathbf U} f_{\tau}(\mathbf L(\mathbf U^{(t)}))$
		\State Update
		\[
		\mathbf U^{(t+1)}
		=
		\mathbf U^{(t)}+\alpha_t \nabla_{\mathbf U}
		f_{\tau}(\mathbf L(\mathbf U^{(t)}))
		\]
		using backtracking line search
		\State Update $\tau\leftarrow \max(\rho\tau,\tau_{\min})$
		\State $t\leftarrow t+1$
		\Until{convergence}
		\State \textcolor{blue}{Output the iterate $\mathbf L(\mathbf U^{(t')})$, $t'\in\{0,\ldots,t\}$, with the largest $\min_{k\in\mathcal S}h_k(\mathbf L(\mathbf U^{(t')}))/\phi_k^2$}
	\end{algorithmic}
\end{algorithm}

\textcolor{blue}{Since the configuration design problem is nonconvex, Algorithm~\ref{alg_pass}, like any local method, cannot guarantee a global optimum.
However, it is initialized at the configuration
at which it is invoked, i.e., with $\mathbf U^{(0)}$ such that
$\mathbf L(\mathbf U^{(0)})$ equals that configuration. Such
$\mathbf U^{(0)}$ exists for every configuration in the relative interior of
the feasible set, since $u_{m,i}=\log(e^{c\,q_{m,i}}-1)$ with any $c>0$
reproduces given slack fractions $q_{m,i}>0$ through \eqref{eq:softplus_map}
and \eqref{eq:slack_fraction}. Moreover, the algorithm returns the
iterate with the largest worst-device value
$\min_{k\in\mathcal S}h_k(\mathbf L)/\phi_k^2$, including its initial
point. Hence, although it optimizes the smoothed
objective internally, the returned configuration never decreases the
objective of \eqref{prob_pass} relative to its initial configuration.}

\vspace{-5pt}
\subsection{Complexity Analysis}

Let $|\mathcal S|$ denote the number of selected devices. The dominant
computational cost of the proposed PASS configuration algorithm comes from
evaluating the effective channels and their gradients.

For each selected device $k$, computing $\mathbf g_k(\mathbf L)$ requires
summing the contributions of all $MN$ pinching antennas. Likewise,
evaluating the derivatives of the smoothed objective requires computing the
corresponding partial derivatives with respect to these antenna positions.
Therefore, one gradient-evaluation step has complexity
\begin{equation}
	\mathcal O(|\mathcal S|MN).
\end{equation}

The reparameterization from $\mathbf U$ to $\mathbf L(\mathbf U)$ involves
only softplus operations, normalization, and cumulative sums across the $N$
antennas on each waveguide, and hence has complexity $\mathcal O(MN)$. This
term is dominated by the gradient evaluation term above.

Consequently, if the configuration optimization requires $I_3$ iterations, the
total complexity of the PASS-configuration block is
\begin{equation}
	\mathcal O
	\big(
	I_3 |\mathcal S|MN
	\big).
\end{equation}

Thus the antenna-configuration update scales linearly with the number of
selected devices, waveguides, and pinching antennas.

\vspace{-5pt}
\subsection{Discussion}
A fundamental characteristic of the PASS configuration problem is that the
effective channels arise from coherent superposition along each waveguide,
which induces strong coupling among all antenna locations in both amplitude
and phase. As a result, the configuration is not naturally separable across
individual antenna locations: local adjustments of individual antennas may not
lead to predictable local improvements, since each update perturbs the global
interference pattern observed by all devices. Existing PASS geometry designs
often rely on sequential element-wise updates or grid-based placement searches
\cite{MIMOPASS,PASSOTA,PASSAirFL,PASSISAC2}. While effective in their respective
settings, such strategies can become inefficient when the configuration block
must be updated repeatedly within an alternating optimization framework.
\textcolor{blue}{Gradient-based placement has also been developed in
\cite{WDMAPASS}, where the waveguide-range constraint is handled by a bounded
reparameterization and the minimum-spacing constraint by a smoothed penalty
whose weight is progressively increased until the constraint is satisfied.}

The formulation also differs from existing PASS placement problems, which
typically optimize communication-centric metrics such as \textcolor{blue}{the
sum rate \cite{WDMAPASS} or the max-min rate \cite{PASSMultiuser}, where
signals intended for other users or groups act as interference}. In
contrast, the objective here is governed by a worst-device aggregation
constraint induced by AirComp, \textcolor{blue}{in which the signals of all
selected devices, combined by a common receive beamformer, jointly constitute
the desired aggregate. This leads to a configuration design driven by
aggregation feasibility rather than by communication rates.}

The proposed HAGO \textcolor{blue}{method} addresses these challenges through coupled
first-order geometry optimization with a smooth bottleneck objective and a
feasible reparameterization. Thus, all pinching locations are updated jointly
while feasibility is maintained by construction \textcolor{blue}{at every
iteration, without constraint penalties}.

\vspace{-7pt}
\subsection{AirPASS Alternating Optimization Framework}

The proposed AirPASS \textcolor{blue}{framework} alternates between the two blocks developed above:
(i) device selection and receive beamforming for fixed PASS configuration, and
(ii) PASS configuration optimization for fixed selected devices and beamformer. The two blocks are tightly coupled: HRMC determines which devices can be aggregated under the current geometry, and HAGO uses that decision to reshape the geometry so that the next device-selection step becomes more favorable.

\textcolor{blue}{However, since HRMC solves a combinatorial problem
heuristically, improving the effective channels of the currently
selected devices does not by itself guarantee that the next HRMC step selects at
least as many devices. AirPASS therefore embeds the alternation in an
incumbent-retention rule based on the worst feasibility score $\psi$
of Section~V-A; recall that $(\mathcal S,\mathbf m,\mathbf L)$
satisfies the aggregation constraints of \eqref{problem_reformulated} if and
only if $\psi(\mathcal S,\mathbf m,\mathbf L)\ge 1$. At iteration $t$, HAGO is
applied to the incumbent $(\mathcal S^{(t)},\mathbf m^{(t)})$ starting from
$\mathbf L^{(t)}$, and HRMC then computes a candidate
$(\tilde{\mathcal S},\tilde{\mathbf m})$ for the new configuration
$\mathbf L^{(t+1)}$. The candidate replaces the incumbent only if it is
lexicographically at least as good in terms of $(|\mathcal S|,\psi)$, i.e.,
if it selects more devices, or equally many devices with a worst feasibility
score that is at least as large; this order is denoted by
$\succeq_{\mathrm{lex}}$. Since the minimum MSE equals
$\varepsilon/\psi(\mathcal S,\mathbf m,\mathbf L)$, the secondary criterion
favors, among selections of equal size, the one with the smaller aggregation
MSE. With
$\psi^{(t)}\triangleq\psi(\mathcal S^{(t)},\mathbf m^{(t)},\mathbf L^{(t)})$,
the algorithm terminates when $|\mathcal S^{(t)}|$ is unchanged and the
relative increase of $\psi^{(t)}$ does not exceed a tolerance
$\epsilon_{\mathrm{AO}}>0$, or after $I_{\mathrm{AO}}^{\max}$ iterations. The
procedure is summarized in Algorithm~\ref{alg_airpass}, and its behavior is
characterized by the following result.}

\begin{algorithm}[t]
	\caption{AirPASS Alternating Optimization Algorithm}
	\label{alg_airpass}
	\begin{algorithmic}[1]
		\color{blue}
		\State Initialize a strictly feasible $\mathbf L^{(0)}$, obtain $(\mathcal S^{(0)},\mathbf m^{(0)})$ via Algorithm~\ref{alg:hybrid}, and set $t=0$
		\Repeat
		\State Obtain $\mathbf L^{(t+1)}$ via Algorithm~\ref{alg_pass} for fixed $(\mathcal S^{(t)},\mathbf m^{(t)})$, initialized at $\mathbf L^{(t)}$
		\State Obtain a candidate $(\tilde{\mathcal S},\tilde{\mathbf m})$ via Algorithm~\ref{alg:hybrid} for fixed $\mathbf L^{(t+1)}$
		\If{the candidate is at least as good as the incumbent under $\mathbf L^{(t+1)}$ with respect to $\succeq_{\mathrm{lex}}$}
		\State $(\mathcal S^{(t+1)},\mathbf m^{(t+1)})\leftarrow(\tilde{\mathcal S},\tilde{\mathbf m})$
		\Else
		\State $(\mathcal S^{(t+1)},\mathbf m^{(t+1)})\leftarrow(\mathcal S^{(t)},\mathbf m^{(t)})$
		\EndIf
		\State $t\leftarrow t+1$
		\Until{$|\mathcal S^{(t)}|=|\mathcal S^{(t-1)}|$ and $\psi^{(t)}-\psi^{(t-1)}\le\epsilon_{\mathrm{AO}}\psi^{(t-1)}$, or $t=I_{\mathrm{AO}}^{\max}$}
		\State \textbf{Output:} $(\mathcal S^{\star},\mathbf m^{\star},\mathbf L^{\star})=(\mathcal S^{(t)},\mathbf m^{(t)},\mathbf L^{(t)})$
	\end{algorithmic}
\end{algorithm}

{\color{blue}
\begin{proposition}\label{prop:ao}
Suppose that $\mathcal S^{(0)}\neq\emptyset$. Then, for every $t$,
(i) $(\mathcal S^{(t)},\mathbf m^{(t)},\mathbf L^{(t)})$ is feasible for
\eqref{problem_reformulated}; (ii) $|\mathcal S^{(t+1)}|\ge|\mathcal S^{(t)}|$,
and $\psi^{(t+1)}\ge\psi^{(t)}$ whenever
$|\mathcal S^{(t+1)}|=|\mathcal S^{(t)}|$; and (iii) $|\mathcal S^{(t)}|$
increases at most $K-|\mathcal S^{(0)}|$ times and $\psi^{(t)}$ is bounded,
so that the tolerance criterion of Algorithm~\ref{alg_airpass} is met after
finitely many iterations for every $\epsilon_{\mathrm{AO}}>0$, even without the
cap $I_{\mathrm{AO}}^{\max}$.
\end{proposition}
\begin{IEEEproof}
Every nonempty set returned by HRMC has the form
$\{k:z_k(\mathbf m;\mathbf L)\ge1\}$ and is therefore feasible. Since
Algorithm~\ref{alg_pass} is initialized at $\mathbf L^{(t)}$ and returns its
best iterate, $\psi(\mathcal S^{(t)},\mathbf m^{(t)},\mathbf L^{(t+1)})
\ge\psi^{(t)}$. Hence, if the incumbent is feasible under
$\mathbf L^{(t)}$, it remains feasible under $\mathbf L^{(t+1)}$, and the
acceptance rule yields
$(|\mathcal S^{(t+1)}|,\psi^{(t+1)})\succeq_{\mathrm{lex}}
(|\mathcal S^{(t)}|,\psi(\mathcal S^{(t)},\mathbf m^{(t)},\mathbf L^{(t+1)}))
\succeq_{\mathrm{lex}}(|\mathcal S^{(t)}|,\psi^{(t)})$, where an empty
candidate is always rejected because $|\mathcal S^{(t)}|\ge1$. Induction from
the feasible initial triple yields (i) and (ii). For (iii),
$|\mathcal S^{(t)}|\le K$ is integer-valued. Moreover, since
$D_{k,m}(\ell_{m,n})\ge a$ and $e^{-\zeta\ell_{m,n}}\le1$, \eqref{eq:channel} gives
$|g_{k,m}(\boldsymbol\ell_m)|\le\xi\gamma_k\sqrt N/a$, so that
$\psi^{(t)}\le\max_{k\in\mathcal K}\Gamma MN\xi^2\gamma_k^2/(a^2\phi_k^2)$.
After the last increase of $|\mathcal S^{(t)}|$, the sequence $\psi^{(t)}$ is
nondecreasing and bounded, hence convergent, and since $\psi^{(t)}\ge1$, its
relative increments vanish. The tolerance criterion is therefore met after
finitely many iterations.
\end{IEEEproof}
}

\textcolor{blue}{Since \eqref{problem_reformulated} is combinatorial and
nonconvex, Proposition~\ref{prop:ao} does not imply global optimality of the
returned solution.}

\vspace{-7pt}
\subsection{Overall Complexity}

Let $I_{\mathrm{AO}}$ denote the number of outer alternating iterations.
Combining the complexity of the two alternating blocks and the \textcolor{blue}{initial} HRMC pass,
the overall worst-case complexity is
\begin{equation}\label{eq:overall_complexity}
	\mathcal O\!\left(
	I_{\mathrm{AO}}
	\left(
	(I_1+I_2)KM^2+I_3|\mathcal S|MN
	\right)
	+
	(I_1+I_2)KM^2
	\right),
\end{equation}
where $I_1$, $I_2$, and $I_3$ denote the numbers of iterations in the
homotopy-reweighted beamforming optimization, feasibility-margin consolidation,
and PASS configuration optimization, respectively, and $|\mathcal S|$ denotes
the selected-set size during the geometry-update block, upper bounded by $K$. Equivalently, this can be written as
$\mathcal O\!\left((I_{\mathrm{AO}}+1)(I_1+I_2)KM^2+
I_{\mathrm{AO}}I_3|\mathcal S|MN\right)$. \textcolor{blue}{The best-iterate
selection in Algorithm~\ref{alg_pass} and the acceptance test in
Algorithm~\ref{alg_airpass} reuse effective channels and feasibility scores
already computed by the two blocks, and therefore do not change this
order. Evaluating the $K$ effective channels for each HRMC call adds $\mathcal O(KMN)$ operations, which is dominated by the HRMC term since $(I_1+I_2)M\ge N$ in practice.}

{\color{blue}
\vspace{-7pt}
\subsection{Operating Timescale and Overhead}

Within a communication round, the selected set, transmit scaling
coefficients, receive beamformer, denoising factor, and PASS configuration are
held fixed, so that all entries of the model update are aggregated with the
same design. The effective channels in \eqref{eq:channel} depend only on the
device locations, the large-scale gains $\{\gamma_k\}$, and the pinching
locations, so the solution of Algorithm~\ref{alg_airpass} remains valid
as long as the device locations and large-scale gains are unchanged. AirPASS
therefore operates on two timescales. On the slow timescale,
Algorithm~\ref{alg_airpass} is executed at the start of training and
re-executed when the device locations or large-scale gains change
appreciably, which, for quasi-static devices, occurs only after many communication rounds.
On the fast timescale, if the effective channels at the current configuration
vary, e.g., due to propagation effects not captured by \eqref{eq:channel}, the selected set
and beamformer can be updated by Algorithm~\ref{alg:hybrid} alone for fixed
$\mathbf L$, without moving the pinching antennas.

For channel acquisition, a
slow-timescale update requires the locations $(x_k,y_k)$ and large-scale gains
$\gamma_k$ of the $K$ devices, from which $\mathbf g_k(\mathbf L)$ and its
derivatives follow for any configuration through \eqref{eq:channel}, whereas a fast-timescale
update requires only the effective channels
$\mathbf g_k(\mathbf L)\in\mathbb C^M$ at the current configuration, which can
be estimated from $K$ orthogonal uplink pilot symbols observed by the $M$ RF
chains. As for any position-based PASS design, the geometry gain relies on the
phases in \eqref{eq:channel}, i.e., on device locations and antenna positions accurate to within a fraction of
$\lambda$; a location mismatch affects only this gain, since the fast-timescale update
restores feasibility for the effective channels estimated at the realized configuration. For optimization, a slow-timescale update incurs the complexity in
\eqref{eq:overall_complexity}, whereas a fast-timescale update incurs only the HRMC complexity
$\mathcal O((I_1+I_2)KM^2)$; neither depends on the model dimension. For
physical reconfiguration, the pinching antennas are moved only on the slow
timescale; if all antennas move simultaneously with actuation speed
$v_{\mathrm{PA}}$, the reconfiguration delay is
$\max_{m,n}|\ell_{m,n}^{\mathrm{new}}-\ell_{m,n}^{\mathrm{old}}|/v_{\mathrm{PA}}$,
whereas implementations that activate preconfigured candidate positions
\cite{WDMAPASS} incur only the corresponding activation time. All
slow-timescale overheads are thus amortized over the communication rounds
between consecutive geometry updates.
}


\vspace{-7pt}
\section{Experiments}

In this section, we evaluate the proposed \emph{AirPASS} framework and
compare it \textcolor{blue}{with co-located MIMO and fixed-position PASS baselines and with}
alternative strategies for device selection and receive beamforming.
\vspace{-5pt}
\subsection{Experimental Setup}

We report results on two standard image-classification datasets, namely
MNIST and CIFAR-10. For MNIST, we use a lightweight convolutional neural
network with two convolutional layers (with 16 and 32 filters,
respectively), each followed by max-pooling, and then two fully connected
layers with 64 hidden units and a 10-way softmax output. For CIFAR-10, we
use a deeper convolutional neural network with three convolutional blocks,
each containing two convolutional layers (with 32, 64, and 128 filters,
respectively), followed by \textcolor{blue}{group} normalization, \textcolor{blue}{rectified linear unit (ReLU)} activation,
max-pooling, and dropout. The network ends with global average pooling, a
dense layer with 128 hidden units, and a 10-way softmax output.
\textcolor{blue}{CIFAR-10 training uses random-crop and horizontal-flip
augmentation.}

\textcolor{blue}{Learning performance is evaluated at the AP on the global
model obtained after the aggregation of each communication round, so that, for
the AirComp-based schemes, the reported accuracy reflects both the device
selection and the aggregation distortion. The test accuracy is the fraction of correctly classified samples in the
standard held-out test set of the corresponding dataset ($10{,}000$ images not
used for training), where the predicted label is the class with the largest
softmax output. It is computed in the same manner for all schemes, including
ideal FedAvg, and the curves show its mean over the Monte Carlo realizations
at each round.}

\textcolor{blue}{The main simulation parameters, including the network
configuration and the swept parameter values of each figure, are summarized in
Table~\ref{tab:sim_params}.} The noise variance is determined by the
prescribed \textcolor{blue}{SNR}, defined as
$\mathrm{SNR} = \frac{P_0}{\sigma^2}$\textcolor{blue}{, i.e., excluding the path loss in \eqref{eq:channel}. The weights are
$\phi_k=|\mathcal D_k|/\bar D$, i.e., dataset sizes are measured in units of
$\bar D=D/K$ (Section~II-C). With $\varepsilon=0.03$, the number of admissible
devices grows from a few to $K$ between about $24$ and $40$~dB
(Fig.~\ref{fig:device_set}), and the learning experiments use $30$~dB.}

\textcolor{blue}{In each Monte Carlo realization, the device locations,
large-scale gains, and data partition are drawn independently and kept fixed
over all rounds; in line with the slow-timescale operation in Section~V-M, the
design of each scheme is computed once before training. The aggregation error
is emulated by adding zero-mean Gaussian noise with variance $\varepsilon/\psi$,
the minimum MSE of the design, to every aggregated normalized entry, i.e., before the division by
$\sum_{k\in\mathcal S}\phi_k=D_{\mathcal S}/\bar D$ and the inverse normalization. The learning rates are selected for ideal
FedAvg on a validation split. Since ideal FedAvg attains only $21\%$ validation
accuracy on CIFAR-10 after $60$ rounds under the pure three-class split, a
fraction $\chi=0.5$ of the samples of each device is drawn from the pooled
training set \cite{SCAFFOLD}, where $0.5$ is the smallest value in $\{0,0.5,0.8,1\}$
for which ideal FedAvg attains $30\%$; this leaves all dataset sizes, weights, and selected sets
unchanged. Results are averaged over $20$ realizations, or over the first $10$
of them for learning, with $95\%$ Student-$t$ confidence intervals shown as shaded areas or
error bars, and final accuracies are averaged over the last three evaluations.}

\begin{table}[t]
	\color{blue}
	\centering
	\caption{Main simulation parameters}
	\label{tab:sim_params}
	\footnotesize
	\setlength{\tabcolsep}{3pt}
	\renewcommand{\arraystretch}{1.0}
	\begin{tabular}{|>{\raggedright\arraybackslash}p{0.375\columnwidth}|>{\raggedright\arraybackslash}p{0.565\columnwidth}|}
		\hline
		\textbf{Parameter} & \textbf{Value} \\
		\hline
		\multicolumn{2}{|l|}{\textit{Network and PASS geometry}} \\
		\hline
		Number of devices, $K$ & $10$ (Figs.~\ref{fig:mnist_comp}, \ref{fig:cifar_comp}, \ref{fig:M}--\ref{fig:SNR}); $50$ (Figs.~\ref{fig:mnist_comp2}, \ref{fig:cifar_comp2}, \ref{fig:device_set}); both (Fig.~\ref{fig:conv}, Tables~\ref{tab:runtime}, \ref{tab:ablation}) \\
		Number of waveguides, $M$ & $5$ ($K=10$); $10$ ($K=50$); $\{2,3,5,8,12\}$ (Fig.~\ref{fig:M}) \\
		Antennas per waveguide, $N$ & $5$ ($K=10$); $10$ ($K=50$); $\{2,4,8,16,40\}$ (Fig.~\ref{fig:N}) \\
		Device position, $(x_k,y_k)$ & $x_k\sim\mathcal{U}[4,40]$~m, $y_k\sim\mathcal{U}[0,W]$, $W=10$~m \\
		Large-scale gain, $\gamma_k$ & $\mathcal{U}[0.8,1.2]$ \\
		Waveguide length, $L_m$ & $40$~m \\
		Waveguide position, $\bar y_m$ & $(m-\frac12)W/M$, i.e., spacing $d=W/M$ (Fig.~\ref{fig:airpass_system_model}) \\
		Waveguide height, $a$ & $4$~m \\
		Minimum spacing, $\Delta\ell$ & $\lambda/2=0.0625$~m \\
		\hline
		\multicolumn{2}{|l|}{\textit{Channel and aggregation}} \\
		\hline
		Carrier wavelength, $\lambda$ & $0.125$~m \\
		Refractive index, $i_{\mathrm{ref}}$ & $1.45$ \\
		In-waveguide attenuation, $\zeta$ & $0$; $0.0092$~m$^{-1}$ (Table~\ref{tab:ablation}) \\
		Free-space constant, $\xi$ & $1$ \\
		Transmit power budget, $P_0$ & $1$ \\
		Aggregation weights, $\phi_k$ & $|\mathcal D_k|/\bar D$, $\bar D=D/K$ \\
		MSE threshold, $\varepsilon$ & $0.03$ \\
		SNR, $P_0/\sigma^2$ & $30$~dB; $\{26,28,30,32,34,40\}$~dB (Fig.~\ref{fig:SNR}); $24$--$40$~dB (Fig.~\ref{fig:device_set}); $\{26,30,34\}$~dB (Table~\ref{tab:ablation}) \\
		\hline
		\multicolumn{2}{|l|}{\textit{Federated learning}} \\
		\hline
		Training/validation samples & Class-balanced: $48{,}000/5{,}000$ (MNIST), $45{,}000/5{,}000$ (CIFAR-10) \\
		Data partition & Three of ten classes per device; CIFAR-10: similarity $\chi=0.5$ \cite{SCAFFOLD} \\
		Local optimizer & Stochastic gradient descent, momentum $0.9$, weight decay $5\times10^{-4}$ (CIFAR-10) \\
		Local learning rate & $0.02$ (MNIST); $0.01$ (CIFAR-10) \\
		Batch size & $32$ (MNIST); $64$ (CIFAR-10) \\
		Local steps per round, $J$ & $10$ \\
		Communication rounds & $100$; $150$ (CIFAR-10, $K=10$) \\
		Monte Carlo realizations & $10$ (learning); $20$ (optimization) \\
		\hline
	\end{tabular}
	\vspace{-10pt}
\end{table}

\textcolor{blue}{Table~\ref{tab:alg_params} summarizes all parameters of the
proposed algorithms with their values, update rules, and stopping criteria,
grouped by block; all line searches use Armijo backtracking. The smoothing parameters
$\beta$, $\mu$, and $\tau$ follow geometric homotopy schedules from initial
values that keep the surrogates smooth to final values that make them tight,
and $\tau_{\mathrm m}$ is fixed at a value that makes the consolidation
soft-min tight: the softplus
approximation \eqref{eq:softplus} exceeds the hinge violation by at most
$\log 2/\beta_{\max}$, and the soft-min objectives \eqref{problem_margin}
and \eqref{eq:lse_hago} underestimate the corresponding exact minima by at
most $\tau_{\mathrm m}\log|\mathcal A|$ and
$\tau_{\min}\log|\mathcal S|$, respectively. The threshold $\delta$ of HRMC and the
subset-cutting weight ($0.6$) of the matching-pursuit benchmark were selected on
separate validation realizations; over $\delta\in\{0.2,0.3,0.4,0.5,0.7\}$, the
fraction of selected devices varied only between $46.0\%$ and $47.2\%$.}

\begin{table}[t]
	\color{blue}
	\centering
	\caption{Parameters of the proposed algorithms}
	\label{tab:alg_params}
	\footnotesize
	\setlength{\tabcolsep}{2.5pt}
	\renewcommand{\arraystretch}{1.0}
	\begin{tabular}{|>{\raggedright\arraybackslash}p{0.10\columnwidth}|>{\raggedright\arraybackslash}p{0.29\columnwidth}|>{\raggedright\arraybackslash}p{0.185\columnwidth}|>{\raggedright\arraybackslash}p{0.33\columnwidth}|}
		\hline
		\textbf{Symbol} & \textbf{Meaning} & \textbf{Value} & \textbf{Update / stopping rule} \\
		\hline
		\multicolumn{4}{|l|}{\textit{HRMC, homotopy-reweighted stage (Algorithm~\ref{alg:hybrid}, lines 4--8)}} \\
		\hline
		$\beta$ & Softplus sharpness in \eqref{eq:softplus} & $10$; $\beta_{\max}=100$ & Increased by \eqref{eq:homotopy_hrmc} \\
		$\kappa_\beta$ & Growth factor of $\beta$ & $1.5$ & Fixed \\
		$\mu$ & Smoothing of the penalty \eqref{eq:sat_penalty} & $0.08$; $\mu_{\min}=0.01$ & Decreased by \eqref{eq:homotopy_hrmc} \\
		$\kappa_\mu$ & Decay factor of $\mu$ & $1.5$ & Fixed \\
		$w_k$ & Reweighting coefficient in \eqref{problem_smooth} & $1$ (initial) & \eqref{eq_weight_update} after each homotopy step \\
		$\epsilon_w$ & Reweighting regularizer in \eqref{eq_weight_update} & $0.05$ & Fixed \\
		-- & Initial RCG step size & $3\times10^{-2}$ & Armijo backtracking \\
		-- & Stopping, inner RCG & -- & Gradient norm reduced $10^{6}$-fold, relative decrease $\le10^{-10}$, or $200$ iterations \\
		-- & Stopping, homotopy loop & -- & $\beta=\beta_{\max}$, $\mu=\mu_{\min}$, and $\bar{\mathcal S}$ unchanged; at most $20$ steps \\
		\hline
		\multicolumn{4}{|l|}{\textit{HRMC, margin consolidation (Algorithm~\ref{alg:hybrid}, lines 10--17)}} \\
		\hline
		$\delta$ & Active-set threshold in \eqref{eq:active_set} & $0.7$ & Fixed \\
		$\tau_{\mathrm m}$ & Soft-min parameter in \eqref{problem_margin} & $0.01$ & Fixed \\
		-- & Stopping, RCG & -- & As inner RCG, at most $300$ iterations \\
		\hline
		\multicolumn{4}{|l|}{\textit{HAGO (Algorithm~\ref{alg_pass})}} \\
		\hline
		$\tau$ & Log-sum-exp smoothing in \eqref{eq:lse_hago} & $\tau_0=0.1/\Gamma$; $\tau_{\min}=0.01/\Gamma$ & Decreased by \eqref{eq:homotopy_hago} \\
		$\rho$ & Decay factor of $\tau$ & $0.8$ & Fixed \\
		$\alpha_t$ & Ascent step size in \eqref{eq:hago_step} & $4\times10^{-2}$ (initial) & Barzilai--Borwein trial, Armijo backtracking \\
		-- & Stopping & -- & $200$ iterations, or relative gain of $\psi$ below $10^{-6}$ over $50$ iterations at $\tau_{\min}$ \\
		\hline
		\multicolumn{4}{|l|}{\textit{AirPASS alternating optimization (Algorithm~\ref{alg_airpass})}} \\
		\hline
		$\epsilon_{\mathrm{AO}}$ & Relative tolerance on $\psi^{(t)}$ & $10^{-3}$ & Stopping rule of Algorithm~\ref{alg_airpass} \\
		$I_{\mathrm{AO}}^{\max}$ & Maximum outer iterations & $50$ & Stopping rule of Algorithm~\ref{alg_airpass} \\
		$\mathbf L^{(0)}$ & Initial configuration & Uniform spacing & -- \\
		\hline
		\multicolumn{4}{|l|}{\textit{All line searches}} \\
		\hline
		-- & Armijo constant; shrink factor & $10^{-4}$; $0.5$ & Every iteration \\
		\hline
	\end{tabular}
	\vspace{-10pt}
\end{table}

\vspace{-7pt}
\subsection{Benchmark Schemes}

We consider the following benchmark schemes\textcolor{blue}{, all of which follow the same local training and FedAvg update of Section~II and differ only in the selected set and in the error of the aggregated update}.

\begin{itemize}
	\item \textbf{FedAvg}: Ideal error-free aggregation \textcolor{blue}{over all devices,} used as an upper bound.

	\item \textbf{AirPASS}: The proposed PASS-enabled AirFL framework with
	HRMC for device selection and receive beamforming, and HAGO for
	pinching-antenna geometry optimization.

	\item \textcolor{blue}{\textbf{Fixed-position PASS--HRMC}: AirPASS without
	HAGO, i.e., the same PASS receiver with uniformly spaced antennas, which
	isolates the gain of geometry optimization under identical hardware.}

	\item \textbf{Co-located MIMO--HRMC ($M$ and $M\times N$)}:
	Conventional centralized fixed-array MIMO baselines with either $M$
	\textcolor{blue}{or} $M\times N$ receive antennas, using the same HRMC block as in AirPASS. The $M$-antenna
	case matches the number of waveguides, whereas the $M\times N$-antenna
	case matches the total number of antenna elements in the PASS receiver.
	The receive antennas form a uniform linear array with inter-element
	spacing $\lambda/2$, centered at the midpoint of the PASS structure,
	and the PASS channel model in Section II-B is replaced by the
	corresponding conventional co-located MIMO uplink channel.
	\textcolor{blue}{Each antenna has its own RF chain.}

	\item \textbf{AirPASS--SDR-DC}: A variant of AirPASS in which the
	HRMC block is replaced by the SDR-DC method in
	\cite{DCSDR}, while the HAGO geometry-optimization block is kept
	unchanged.

	\item \textbf{AirPASS--MP}: A variant of AirPASS in which the HRMC block is replaced by the matching-pursuit \textcolor{blue}{(MP)} method
	in \cite{DeviceSchedulingOTAFL}, while the HAGO
	geometry-optimization block is kept unchanged.
\end{itemize}

\begin{figure*}[t]
	\centering
	\begin{minipage}[t]{0.32\textwidth}
		\centering
		\includegraphics[width=\linewidth]{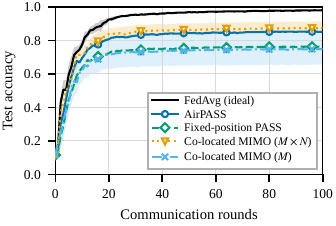}
		\caption{Test accuracy versus communication rounds on MNIST for AirPASS,
		ideal FedAvg, \textcolor{blue}{fixed-position PASS,} and conventional co-located MIMO baselines. $K = 10, M = 5, N = 5$, SNR = \textcolor{blue}{30} dB.}
		\label{fig:mnist_comp}
	\end{minipage}
\hfill
	\begin{minipage}[t]{0.32\textwidth}
		\centering
		\includegraphics[width=\linewidth]{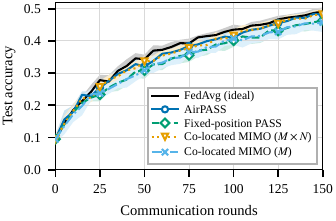}
		\caption{Test accuracy versus communication rounds on CIFAR-10 for AirPASS,
		ideal FedAvg, \textcolor{blue}{fixed-position PASS,} and conventional co-located MIMO baselines. $K = 10, M = 5, N = 5$, SNR = \textcolor{blue}{30} dB.}
		\label{fig:cifar_comp}
	\end{minipage}
\hfill
	\begin{minipage}[t]{0.32\textwidth}
		\centering
		\includegraphics[width=\linewidth]{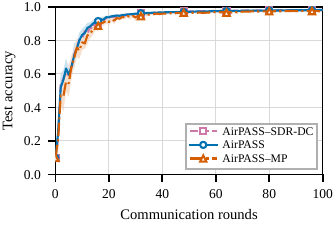}
		\caption{Comparison of strategies for device selection and receive beamforming within PASS-enabled
		AirFL on MNIST. $K = 50, M =10, N = 10$, SNR = \textcolor{blue}{30} dB.}
		\label{fig:mnist_comp2}
	\end{minipage}
	\vspace{-12pt}
\end{figure*}

\begin{figure*}[t]
	\centering
	\begin{minipage}[t]{0.32\textwidth}
		\centering
		\includegraphics[width=\linewidth]{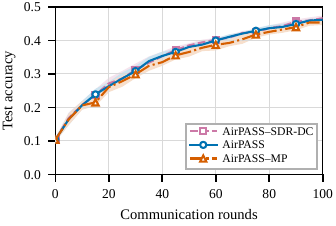}
		\caption{Comparison of strategies for device selection and receive beamforming within PASS-enabled
		AirFL on CIFAR-10. $K = 50, M = 10, N = 10$, SNR = \textcolor{blue}{30} dB.}
		\label{fig:cifar_comp2}
	\end{minipage}
\hfill
	\begin{minipage}[t]{0.32\textwidth}
		\centering
		\includegraphics[width=\linewidth]{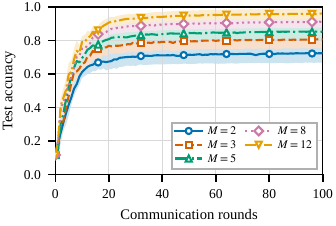}
		\caption{Impact of the number of waveguides $M$ on the \textcolor{blue}{test accuracy} of
		AirPASS on MNIST. $K = 10, N = 5$, SNR = \textcolor{blue}{30} dB.}
		\label{fig:M}
	\end{minipage}
\hfill
	\begin{minipage}[t]{0.32\textwidth}
		\centering
		\includegraphics[width=\linewidth]{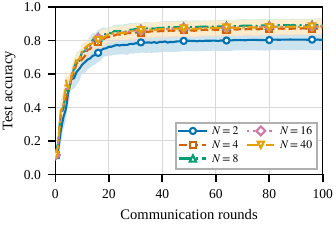}
		\caption{Impact of the number of pinching antennas per waveguide $N$ on the
		\textcolor{blue}{test accuracy} of AirPASS on MNIST. $K = 10, M = 5$, SNR = \textcolor{blue}{30} dB.}
		\label{fig:N}
	\end{minipage}
	\vspace{-12pt}
\end{figure*}

\begin{figure*}[t]
	\centering
	\begin{minipage}[t]{0.32\textwidth}
		\centering
		\includegraphics[width=\linewidth]{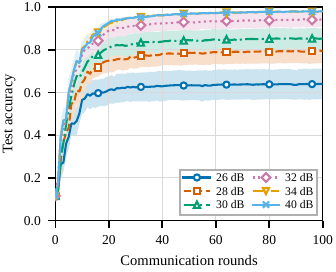}
		\caption{Impact of SNR on the \textcolor{blue}{test accuracy} of AirPASS on MNIST. $K = 10, M = 5, N = 5$.}
		\label{fig:SNR}
	\end{minipage}
\hfill
	\begin{minipage}[t]{0.32\textwidth}
		\centering
		\includegraphics[width=\linewidth]{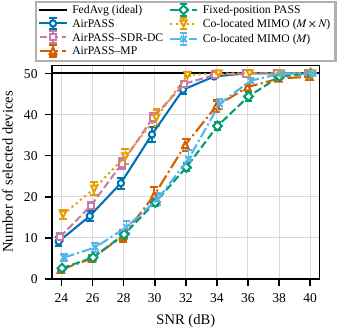}
		\caption{Average number of selected devices for AirPASS and all the benchmark schemes. $K = 50, M = 10, N = 10$.}
		\label{fig:device_set}
	\end{minipage}
\hfill
	\begin{minipage}[t]{0.32\textwidth}
		\centering
		\includegraphics[width=\linewidth]{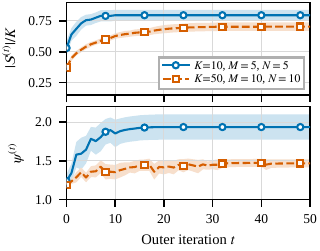}
		\caption{\textcolor{blue}{Fraction of selected devices and worst feasibility
		score over the iterations of Algorithm~\ref{alg_airpass}. SNR $=30$~dB.}}
		\label{fig:conv}
	\end{minipage}
	\vspace{-12pt}
\end{figure*}

\begin{table}[t]
	\color{blue}
	\centering
	\caption{Measured runtime and iteration counts (mean over $10$ separate realizations, SNR $=30$~\textnormal{dB})}
	\label{tab:runtime}
	\footnotesize
	\setlength{\tabcolsep}{2.2pt}
	\renewcommand{\arraystretch}{1.0}
	\begin{tabular}{|l|c|c|c|c|c|c|}
		\hline
		& \multicolumn{3}{c|}{$K=10$, $M=N=5$} & \multicolumn{3}{c|}{$K=50$, $M=N=10$}\\
		\hline
		Method & Time [s] & Iter. & $|\mathcal S|$ & Time [s] & Iter. & $|\mathcal S|$ \\
		\hline
		HRMC & $0.008$ & $139/120$ & $5.3$ & $0.013$ & $184/260$ & $17.6$ \\
		SDR-DC & $0.096$ & $39.6$ & $5.9$ & $4.65$ & $178$ & $20.3$ \\
		Matching pursuit & $0.0002$ & $7.3$ & $3.8$ & $0.0015$ & $38.4$ & $12.7$ \\
		HAGO (one call) & $0.019$ & $200$ & -- & $0.034$ & $200$ & -- \\
		\hline
		AirPASS & $0.29$ & $11.0$ & $7.6$ & $1.29$ & $21.6$ & $33.5$ \\
		AirPASS--SDR-DC & $1.08$ & $10.4$ & $8.0$ & $67.3$ & $30.6$ & $40.7$ \\
		AirPASS--MP & $0.22$ & $11.3$ & $5.2$ & $1.13$ & $29.0$ & $19.9$ \\
		\hline
	\end{tabular}
	\\[2pt]
	\parbox{\columnwidth}{\scriptsize Upper part: selection for the uniform configuration and one HAGO call. Iterations: $I_1/I_2$ (HRMC), semidefinite programs (SDR-DC), selection steps (matching pursuit), $I_3$ (HAGO), and $I_{\mathrm{AO}}$ (lower part).}
	\vspace{-10pt}
\end{table}

\begin{table}[t]
	\color{blue}
	\centering
	\caption{Average number of selected devices for fixed placements, ablated variants of AirPASS, and in-waveguide attenuation (mean $\pm$ 95\% confidence interval over $20$ realizations)}
	\label{tab:ablation}
	\footnotesize
	\setlength{\tabcolsep}{2.2pt}
	\renewcommand{\arraystretch}{1.0}
	\begin{tabular}{|l|c|c|c|c|}
		\hline
		& $K=10$ & \multicolumn{3}{c|}{$K=50$, $M=N=10$} \\
		Scheme & $30$~dB & $26$~dB & $30$~dB & $34$~dB \\
		\hline
		\multicolumn{5}{|l|}{\textit{HRMC with fixed placement}}\\
		\hline
		Uniform & $5.30\pm0.59$ & $5.2\pm0.5$ & $18.6\pm0.9$ & $37.2\pm1.1$ \\
		Random & $5.00\pm0.55$ & $5.4\pm0.4$ & $16.4\pm0.9$ & $34.8\pm1.0$ \\
		User-aligned & $5.85\pm0.41$ & $5.6\pm0.5$ & $18.2\pm1.0$ & $36.3\pm1.1$ \\
		\hline
		\multicolumn{5}{|l|}{\textit{AirPASS and ablated variants}}\\
		\hline
		AirPASS & $7.95\pm0.44$ & $15.3\pm1.3$ & $35.2\pm1.7$ & $49.3\pm0.4$ \\
		User-aligned init. & $8.35\pm0.35$ & $15.4\pm0.8$ & $36.0\pm1.9$ & $49.7\pm0.2$ \\
		W/o homotopy & $7.65\pm0.59$ & $10.1\pm1.6$ & $33.9\pm2.4$ & $48.9\pm0.5$ \\
		W/o reweighting & $7.95\pm0.44$ & $15.5\pm0.7$ & $37.0\pm1.8$ & $49.0\pm0.7$ \\
		W/o consolidation & $7.65\pm0.35$ & $13.9\pm1.0$ & $32.6\pm1.3$ & $48.7\pm0.6$ \\
		\hline
		\multicolumn{5}{|l|}{\textit{In-waveguide attenuation, $\zeta=0.0092$~m$^{-1}$}}\\
		\hline
		Designed for $\zeta=0$ & -- & $12.5\pm0.7$ & $26.8\pm1.7$ & $49.0\pm0.7$ \\
		Attenuation-aware & -- & $12.8\pm0.7$ & $28.5\pm1.5$ & $49.3\pm0.4$ \\
		\hline
	\end{tabular}
	\vspace{-10pt}
\end{table}

\vspace{-5pt}
\subsection{AirPASS versus Conventional MIMO Baselines}

\textcolor{blue}{Figs.~\ref{fig:mnist_comp} and \ref{fig:cifar_comp} compare
AirPASS with ideal FedAvg, the fixed-position PASS receiver, and the two
co-located baselines. With the same number of RF chains, AirPASS outperforms
the fixed-position PASS receiver and the $M$-antenna array: it
admits $7.7$ devices on average in these experiments instead of $5.4$ and $5.5$, and it raises the
final test accuracy from $76.3\%$ and $74.9\%$ to $85.2\%$ on MNIST and from
$45.8\%$ and $45.5\%$ to $47.9\%$ on CIFAR-10, gains that are significant in
paired comparisons. The
$M\times N$-antenna array, with $N=5$ times more RF chains, admits $8.0$ devices
and attains $87.5\%$ and $47.7\%$, within the confidence intervals of AirPASS.}

\textcolor{blue}{The gap to ideal FedAvg depends on the data partition. Since
the selected set is fixed during training, classes held only by excluded
devices are not learned under the pure three-class split of MNIST, and the
accuracy follows the number of covered classes, on average $8.7$ of $10$ for
AirPASS; in the realization in which all classes are covered, AirPASS attains
$98.1\%$, compared with $98.2\%$ for ideal FedAvg. On CIFAR-10, where the selected
devices hold samples of all classes, AirPASS remains within $0.7$ percentage
points of ideal FedAvg ($48.6\%$).}
\vspace{-5pt}
\subsection{Comparison with Alternative Fixed-Configuration Strategies}

\textcolor{blue}{Figs.~\ref{fig:mnist_comp2} and \ref{fig:cifar_comp2} isolate
the contribution of the HRMC block. With $K=50$, AirPASS--SDR-DC, AirPASS, and
AirPASS--MP admit $38.2$, $34.2$, and $19.7$ devices on average in these experiments, which cover all
classes, and attain $98.2\%$, $98.2\%$, and $97.5\%$ on MNIST and $46.1\%$,
$45.7\%$, and $44.9\%$ on CIFAR-10, respectively. AirPASS thus lies within the
confidence interval of AirPASS--SDR-DC, whereas the lower participation of
AirPASS--MP costs $0.6\pm0.4$ and $0.8\pm0.8$ percentage points on MNIST and
CIFAR-10 in paired comparisons, the latter not significant.}
\vspace{-5pt}
\subsection{\protect\textcolor{blue}{Impact of $M$, $N$, and SNR}}

\textcolor{blue}{Figs.~\ref{fig:M}--\ref{fig:SNR} show the test accuracy of
AirPASS on MNIST. Increasing $M$ from $2$
to $12$ raises the accuracy from $72.3\%$ to $95.9\%$ and the number of selected
devices from $5.1$ to $9.8$, since every waveguide adds a receive dimension.
Increasing $N$ from $2$ to $8$ raises the accuracy from $80.5\%$ to $89.3\%$,
whereas the values for $N\ge8$ lie within each other's confidence intervals,
since additional pinching antennas refine the geometry but add no receive
dimension. The SNR scales all feasibility scores through $\Gamma$: the accuracy
increases from $63.9\%$ at $26$~dB to $97.9\%$ at $34$~dB, where all devices are
admitted and AirPASS matches ideal FedAvg despite the aggregation noise.}

\vspace{-7pt}
\subsection{Device Selection Performance versus SNR}\label{sec:exp_snr}

Fig.~\ref{fig:device_set} shows the average number of selected devices versus SNR for AirPASS and \textcolor{blue}{the} benchmark schemes.
\textcolor{blue}{In terms of the equivalent SNR gain, i.e., the horizontal
distance between the curves over $24$--$32$~dB, AirPASS outperforms the
fixed-position PASS receiver and the $M$-antenna array, which have the same
number of RF chains, by $3.1$--$4.7$~dB and $2.6$--$3.2$~dB, e.g., admitting $35.2$
instead of $18.6$ and $19.4$ devices at $30$~dB. Since the fixed-position PASS receiver even trails
the $M$-antenna array by $0.2$--$2.0$~dB, the $40$-m aperture of PASS yields no gain
without geometry optimization. The $M\times N$-antenna array, with $N=10$ times more RF chains,
is $0.7$--$2.1$~dB ahead of AirPASS, and AirPASS--SDR-DC only $0.3$--$0.9$~dB,
which still corresponds to up to $4.8$ more devices because the curves are steep.
AirPASS--MP is $2.5$--$3.7$~dB behind AirPASS and does not reach full
participation even at $40$~dB.}

\vspace{-7pt}
\subsection{\protect\textcolor{blue}{Convergence and Computational Cost}}\label{sec:exp_runtime}

\textcolor{blue}{Fig.~\ref{fig:conv} shows $|\mathcal S^{(t)}|/K$ and
$\psi^{(t)}$ over the outer iterations of Algorithm~\ref{alg_airpass}, averaged
over $20$ realizations per $K$. In line with Proposition~\ref{prop:ao}, all $40$
trajectories are lexicographically nondecreasing in
$(|\mathcal S^{(t)}|,\psi^{(t)})$. Starting from the uniform configuration, the
number of selected devices grows from $5.30$ to $7.95$ for $K=10$ and from
$18.6$ to $35.2$ for $K=50$, with $90\%$ of the average gain reached after $6$
and $18$ iterations, respectively. All runs except one with $K=50$, which stops
at $I_{\mathrm{AO}}^{\max}$, terminate by the tolerance criterion, and the
average numbers of outer iterations are $12.9$ and $28.1$, respectively.}

\textcolor{blue}{Table~\ref{tab:runtime} reports runtimes measured in a single
Python process (Python 3.13, NumPy 2.3) on a laptop with a $12$-core Apple-silicon processor, with the semidefinite programs of SDR-DC solved
by CLARABEL in CVXPY 1.8. For a fixed configuration, HRMC takes $8$ and $13$~ms for
$K=10$ and $50$, whereas SDR-DC takes $0.10$ and $4.6$~s, as its number of
semidefinite programs grows from $40$ to $178$. The complete AirPASS algorithm
takes $0.29$ and $1.3$~s, compared with $1.1$ and $67$~s for AirPASS--SDR-DC,
i.e., about $2\%$ of its runtime for $K=50$ at comparable accuracy (Section~VI-D),
consistent with Table~\ref{tab:complexity_comparison}. This cost is incurred once per
slow-timescale update (Section~V-M) and thus amortized over many rounds.}

\vspace{-7pt}
\subsection{\protect\textcolor{blue}{Geometry Optimization, Ablation, and Attenuation}}\label{sec:exp_ablation}

\textcolor{blue}{The upper part of Table~\ref{tab:ablation} combines HRMC with
fixed uniform, random (uniform over the feasible configurations), and user-aligned placements, the latter placing the
antennas of each waveguide above the devices laterally closest to it. For
$K=50$, AirPASS roughly doubles their participation at $30$~dB and nearly triples it at
$26$~dB, even when initialized at the user-aligned placement. Relative to uniform spacing,
the user-aligned placement and AirPASS raise the channel gains
$\|\mathbf g_k\|^2$ by only $0.6$ and $1.0$~dB on average for $K=50$, yet only
AirPASS raises the participation, by $3.1$--$4.7$~dB in SNR (Section~\ref{sec:exp_snr}).
HAGO thus mainly aligns the effective channels for a common beamformer, which
requires wavelength-scale control of the antenna positions. To gauge its optimality, we solve $100$ small instances of \eqref{prob_pass} per setting
($|\mathcal S|=3$, $5$-m waveguides, $(M,N)\in\{(1,2),(2,1)\}$)
exhaustively on a $1$-mm grid with local refinement. HAGO attains on average
$88\%$ of the optimal objective, compared with $21$--$41\%$ for the fixed
placements, and running it to convergence raises this ratio by at most $1.3$
percentage points.}

\textcolor{blue}{In the middle part, the variants without homotopy, reweighting, and
consolidation start HRMC at $(\beta_{\max},\mu_{\min})$, fix $w_k=1$, and stop
HRMC after its first stage, respectively. Removing HAGO, i.e., the uniform
placement in the upper part, causes by far the largest loss, whereas the homotopy adds about $50\%$ to the
participation at $26$~dB, the margin consolidation up to $2.6$ devices, and the
reweighting no significant change. In the lower part and over $24$--$40$~dB, a placement designed for
$\zeta=0$ admits at most $1.8$ fewer devices on the attenuated channel than its
attenuation-aware refinement by Algorithm~\ref{alg_airpass}. This confirms that neglecting the
attenuation in the design is nearly harmless
\cite{PASSAttenuation,PASSAttenuationBlockage}, whereas the attenuation itself
shifts the participation-versus-SNR curve by $0.8$--$1.1$~dB over $26$--$32$~dB, i.e., the
lossless baseline is optimistic by about $1$~dB.}

\section{Conclusion}

This paper studied over-the-air federated learning with pinching antenna
systems and proposed \emph{AirPASS}, a PASS-enabled framework that
jointly optimizes device selection, receive beamforming, and pinching-antenna
locations under the learning-oriented objective of maximizing the
number of participating devices subject to an aggregation-distortion
constraint. For the joint device-selection and receive-beamforming block, we developed a
homotopy-Riemannian margin-consolidation algorithm that combines smooth
cardinality approximation, homotopy-reweighted Riemannian optimization, and
active-set feasibility-margin consolidation. For the geometry-design block, we
proposed a homotopy-assisted
gradient-based PASS optimization method with a feasible reparameterization
that enforces antenna ordering and minimum spacing by construction.
\textcolor{blue}{Simulation results showed that, with the same number of RF chains, AirPASS substantially increases the number of admissible devices and improves the learning performance over co-located MIMO and fixed-position PASS receivers, mainly owing to the geometry optimization, and attains a learning accuracy comparable to that of a co-located array with one RF chain per pinching antenna. Moreover, AirPASS matches the participation of SDR-DC to within $1$~dB in SNR, with comparable accuracy, at about $2\%$ of its runtime for $50$ devices, and it clearly outperforms matching-pursuit scheduling.}


\newpage

\end{document}